\documentclass[12pt]{article}
\usepackage{amsmath,bbold,amsfonts,amssymb,picture,graphicx,xcolor,soul,hyperref,cleveref,todonotes,
comment,subcaption,booktabs,bm}
\usepackage[toc,title,page]{appendix}
\usepackage[font=small,labelfont=bf]{caption}
\allowdisplaybreaks

\begin{document}
\thispagestyle{empty}

\def\theequation{\arabic{section}.\arabic{equation}}
\def\a{\alpha}
\def\b{\beta}
\def\g{\gamma}
\def\d{\delta}
\def\dd{\rm d}
\def\e{\epsilon}
\newcommand{\q}{\mathsf{q}}
\def\S{\mathsf{S}}
\def\ve{\varepsilon}
\def\z{\zeta}
\def\teta{\tilde\eta}
\def\uqsu2{\mathcal{U}_{\mathsf{q}}(\mathfrak{su}_2)}

\newcommand{\h}{\hspace{0.5cm}}

\begin{titlepage}
\vspace*{-1.cm}
\renewcommand{\thefootnote}{\fnsymbol{footnote}}
\begin{center}
{\Large \bf Hagedorn singularity in exact $\uqsu2$ $S$-matrix theories with arbitrary spins}
\end{center}
\vskip .8cm \centerline{\bf Changrim  Ahn$^1$, Tommaso Franzini$^2$, and Francesco Ravanini$^{3,4}$ }

\vskip 10mm

\centerline{\sl $^1$Department of Physics} \centerline{\sl Ewha Womans University}
 \centerline{\sl 52, Ewhayeodae-gil, Seoul 03760, S. Korea}

\vskip 0.8cm

\centerline{\sl $^2$Department of Physics, Astronomy and Mathematics}
\centerline{\sl University of Hertfordshire} 
\centerline{\sl Hatfield, AL10 9AB, United Kingdom}

\vskip 0.8cm

\centerline{\sl $^3$Department of Physics and Astronomy}
\centerline{\sl University of Bologna}
\centerline{\sl Via Irnerio 46, 40126 Bologna, Italy}

\vskip 0.3cm

\centerline{\sl $^4$ Istituto Nazionale di Fisica Nucleare, Sezione di Bologna}
\centerline{\sl Via Irnerio 46, 40126 Bologna, Italy}

\vskip 10mm

\baselineskip 18pt

\begin{center}
{\bf Abstract}
\end{center}
Generalizing the quantum sine-Gordon and sausage models, we construct exact $S$-matrices for higher spin representations with quantum $\uqsu2$ symmetry, which satisfy unitarity, crossing-symmetry and the Yang-Baxter equations with minimality assumption, i.e. without any unnecessary CDD factor. The deformation parameter $\q$ is related to a coupling constant. Based on these $S$-matrices, we derive the thermodynamic Bethe ansatz equations for $\q$ a root of unity in terms of a universal kernel where the nodes are connected by graphs of non-Dynkin type. We solve these equations numerically to find out Hagedorn-like singularity in the free energies at some critical scales and find a universality in the critical exponents, all near 0.5 for different values of the spin and the coupling constant.

\end{titlepage}
\newpage
\baselineskip 18pt

\def\nn{\nonumber}
\def\tr{{\rm tr}\,}
\def\p{\partial}
\newcommand{\non}{\nonumber}
\newcommand{\bea}{\begin{eqnarray}}
\newcommand{\eea}{\end{eqnarray}}
\newcommand{\bde}{{\bf e}}
\renewcommand{\thefootnote}{\fnsymbol{footnote}}
\newcommand{\be}{\begin{eqnarray}}
\newcommand{\ee}{\end{eqnarray}}
\newcommand{\beq}{\begin{equation}}
\newcommand{\eeq}{\end{equation}}
\newcommand{\ttbar}{$T{\overline T}$}

\vskip 0cm

\renewcommand{\thefootnote}{\arabic{footnote}}
\setcounter{footnote}{0}

\setcounter{equation}{0}

\section{Introduction}

Integrable quantum field theories (QFTs) in two dimensions are valuable models for understanding various non-perturbative properties.
Thanks to an infinite number of conserved charges, 
multi-particle scattering processes are purely elastic so that the number and momenta of the asymptotic particles are preserved and their amplitudes are factorized into a product of two-particle $S$-matrices. 
These $S$-matrices are often determined completely by Yang-Baxter equations (YBEs) along with unitarity and crossing symmetry, and other analytic {\em bootstrap} constraints such as bound state poles in the physical strip if they exist.

Consider the sine-Gordon QFT as an example. Although it is defined by a Lagrangian of a self-interacting scalar field, there is a regime of its coupling constant (the so-called repulsive regime) where the asymptotic particles are not  bosonic, but a doublet of a soliton - antisoliton pair, whose $S$-matrices have been obtained in \cite{zamzam}
by the bootstrap method and YBE.
The $S$-matrix is proportional to the $\mathcal{R}$ matrix of the quantum group
$\uqsu2$ \cite{BerLeC} where the soliton pair belongs to a fundamental spin $1/2$ representation and the deformation parameter $\q$ is related to the coupling constant.
The $S$-matrix satisfies the YBE as the $\mathcal{R}$ matrix does by construction.
An overall scalar function $S_0(\theta)$, which cannot be fixed by the YBE, is determined by unitarity, crossing symmetry, and additional analytic properties.
Instead, when the coupling constant is in the attractive regime, the overall scalar function 
should have additional factors with poles in the physical strip $0<\Im m\ \theta<\pi$ which correspond to ``breathers'', the soliton-antisoliton bound states.
Strictly speaking, the $S$-matrix bootstrap method does not completely fix the scalar function since one can
always multiply the so-called ``CDD factors'' which satisfy all the $S$-matrix axioms but without new bound state poles in the physical strip.
It is customary to adopt a ``minimality'' assumption, which amounts to include no such CDD factors.

The $S$-matrices constructed in this way can provide computational tools for integrable QFTs even when conventional Lagrangians are either not given or given in terms of scale-dependent effective parameters.
Although the $S$-matrices are on-shell quantities,  they can be used to compute finite-size effects using the Thermodynamic Bethe ansatz (TBA) \cite{alyosha_tba} method and even off-shell quantities like correlation functions in terms of the so-called form factor expansions.
The ``inverse scattering program'' is a philosophical framework in which the exact $S$-matrices play a fundamental role. 
This point of view has been successfully applied to many theories in 2D relativistic integrable QFTs 
such as the sausage model, a deformed non-linear sigma model on an effective background metric \cite{foz}, as well as higher dimensional gauge theories related to AdS/CFT duality \cite{ads/cft}.

Notwithstanding, we want to raise a question on the inverse scattering program: can there be a scale where the 
$S$-matrices are not valid anymore? 
In QFTs, it is not uncommon that the fundamental degrees of freedom change depending on the scale.
For example, quarks replace color singlet hadrons as fundamental degrees of freedom when the temperature scale goes beyond $\Lambda_{\rm QCD}$ in the strong interactions, which is the so-called Hagedorn phase transition \cite{Hagedorn}.
A similar transition is also predicted in string theories \cite{stringHag}.
Even for integrable QFTs, it is not excluded in principle that the particles may either change into others or even lose their validity as asymptotic states at a certain energy scale.
This possibility may conflict apparently with the inverse scattering program and its tools such as TBA which assume the 
$S$-matrices are valid at all scales. 
If this happens, such fundamental building blocks of integrability as symmetry, asymptotic particle spectrum, and their $S$-matrices will all be in trouble. 

Recent developments on \ttbar\ deformations have addressed this question. 
If an integrable QFT is deformed by irrelevant operators built from the energy-momentum tensor components $T$ and $\bar{T}$ and their descendants, $T_s$ and $\bar{T}_s$, the $S$-matrix gets extra CDD factors, which in turn lead to Hagedorn-like singularity in the free energies \cite{ttbar1,ttbar2,ttbar3,ttbar4,ttbar5}.
In exceptional cases where these \ttbar\ deformations are fine-tuned, the deformed theories and their particle spectrum can avoid the singularity and reach UV complete theories, which interpolate between UV and IR conformal QFTs by exact RG flows \cite{AhnLeC}.
But general irrelevant deformations should lead to the singularity 
at a certain scale.
Even though the \ttbar\ deformations provide interesting examples of Hagedorn singularity,
one cannot exclude that other irrelevant operators may play a central role in the singular behaviors of  the Hagedorn-like singularity.
It is not easy, within our scattering framework, to identify inherent features of the original scattering theories which trigger the singularity.
To understand this, it will be useful to find QFTs that show the singularity {\it without any additional irrelevant deformations}.

In this paper, we will address this question based on a new class of exact integrable factorized scattering theories. 
The integrable QFTs we are considering here are a generalization of the sine-Gordon and the sausage models to the higher spin representation of the quantum $\uqsu2$ group. 
The deformation parameter $\q$ is related to a coupling constant.
A rational version of this scattering theory for $\q\to 1$ (zero coupling limit) has been explored before in \cite{AladimMartins}. Similar $\mathcal{R}$-matrix-related constructions for higher spin representations appear in the context of statistical lattice models like the integrable XXZ higher spin chains introduced by a fusion method in 
\cite{Kir-Resh1}.

In \cref{sec2}, we derive exact $S$-matrices for spin $s$ multiplets of quantum $\uqsu2$ group and solve the $S$-matrix bootstrap conditions exactly.
We emphasize that our $S$-matrices are ``minimal", in the sense that they do not have any additional CDD factor, since it makes the UV behavior more singular.
As far as we know, these $S$-matrices are the first ones for higher spin particles with non-trivial interactions. 
We will derive the TBA equations rigorously in the form of a universal kernel and use them to study the free energy using the $S$-matrices in \cref{sec3} and \cref{appA}. 
The TBA systems are non-linear integral equations that normally cannot be solved analytically except for the UV and IR limits.
We need, therefore in \cref{sec4}, to rely on numerical analysis of the TBA to find the critical scale where the Hagedorn singularities occur and numerically estimate the critical exponents.
We find that while the critical scales depend both on the values of spin and on the coupling constants,
the critical exponents are very close to $1/2$ for all spins and couplings, which suggests a universal behavior with square root singularity.  
We will summarize and discuss some open questions in the concluding \cref{sec5}.
An attempt to understand the analytic mechanisms underlying the Hagedorn-like singularity in the TBA systems is reported in \cref{appB}, where we analyze a simpler TBA system by using a toy kernel to find out exact critical temperature.

\setcounter{equation}{0}
\section{Exact $S$-matrix for particles with an arbitrary spin}\label{sec2}

We first consider a completely factorized scattering theory between particles that belong to spin $s$ irreducible representation of the $\mathfrak{su}_2$ algebra generated by $\mathbb{J}_{\pm}$, $\mathbb{J}_3$ obeying the commutation relations
\beq
[\mathbb{J}_{\pm},\mathbb{J}_3]=\pm\mathbb{ J}_{\pm}\quad,\quad[\mathbb{J}_+,\mathbb{J}_-]=2\mathbb{J}_3
\eeq
and having Casimir operator $\mathbb{Q}=\mathbb{J}^2=(\mathbb{J}_+\mathbb{J}_-+\mathbb{J}_- \mathbb{J}_+)/2+\mathbb{J}_3^2$.
The spin $s$ are non-negative integers or half-integers for finite-dimensional irreducible
representations.
The on-shell particles can be denoted by $A_m(\theta)$ where $\theta$ is the rapidity which parametrizes
the energy-momentum $E=\mathsf{m}\cosh\theta,\ p=\mathsf{m}\sinh\theta$ and
$m$ is the magnetic quantum number of the $\mathfrak{su}_2$ algebra with $m=s,s-1,\ldots,-s$.
\footnote{Please note the difference between the symbol $\mathsf{m}$ (mass) and $m$, the magnetic
$\mathfrak{su}_2$ quantum number.}
If the scattering theory is integrable, multi-particle scattering amplitudes are decomposed into
two-particle elastic $S$-matrix element
\bea
S_{m_1 m_2}^{m'_1 m'_2}(\theta_1-\theta_2):
A_{m_1}(\theta_1)A_{m_2}(\theta_2)\quad\to\quad
A_{m'_2}(\theta_2)A_{m'_1}(\theta_1).
\label{defS}
\eea
This $S$-matrix satisfies the Yang-Baxter equations ($\theta_{ij}\equiv\theta_i-\theta_j$)
\beq
S_{12}(\theta_{12})S_{13}(\theta_{13})S_{23}(\theta_{23})=
S_{23}(\theta_{23})S_{13}(\theta_{13})S_{12}(\theta_{12})
\eeq
along with unitarity 
\beq
\sum_{n_1,n_2}S_{m_1 m_2}^{n_1,n_2}(\theta)S^{m'_1 m'_2}_{n_1,n_2}(-\theta)=
\delta_{m_1}^{m'_1}\,\delta_{m_2}^{m'_2}
\eeq
and crossing symmetry \eqref{crossing}.

We can decompose this two-particle $S$-matrix into projectors
\bea
\label{undef}
S(\theta)=\mathsf{P}\,\sum_{J=0}^{2s}f^{[J]}(\theta)\,\mathbb{P}^{[J]},
\eea
where $\mathsf{P}$ is the permutation matrix and $\mathbb{P}^{[J]}$ the projector into the spin-$J$ 
representation, 
\bea
\mathbb{P}^{[J]}=\sum_{M=-J}^{J}\vert J,M\rangle\langle J,M\vert,
\eea
which satisfies
\bea
\sum_{J=0}^{2s}\mathbb{P}^{[J]}=\mathbb{I},\qquad
{\rm and}\qquad \left(\mathbb{P}^{[J]}\right)^2=\mathbb{P}^{[J]}.
\eea
Their matrix elements are written in terms of the Clebsch-Gordan coefficients
\bea
{\mathbb{P}^{[J]}}_{m_1 m_2}^{m'_1 m'_2}=\sum_{M=-J}^{J}\langle s,m'_1;s,m'_2\vert J,M\rangle
\langle J,M\vert s,m_1;s,m_2\rangle.
\eea

The Yang-Baxter equation determines the scalar functions 
\bea
f^{[J]}(\theta)=
\prod_{k=1}^{J}\frac{i\pi k-\theta}{i\pi k+\theta},
\eea
up to an overall function which can be fixed by unitarity and crossing symmetry.
This $S$-matrix has been studied in \cite{AladimMartins}.

We extend this ``rational''  $S$-matrix to the ``trigonometric''  one $\S$ by introducing certain interactions
in terms of a coupling constant which is related to a deformation parameter ${\q}\in\mathbb{C}$ of the quantum group symmetry algebra $\uqsu2$, generated by $\mathbb{J}_{\pm},\q^{\pm\mathbb{J}_3}$ such that
\beq
[\mathbb{J}_{\pm},\mathbb{J}_3]=\pm\mathbb{ J}_{\pm}\quad,\quad[\mathbb{J}_+,\mathbb{J}_-]=[2\mathbb{J}_3]
\eeq
and with Casimir operator
\beq
\mathbb{Q}=\mathbb{J}_+\mathbb{J}_- + \left[\mathbb{J}_3 -\frac12\right]=\mathbb{J}_-\mathbb{J}_+ + \left[\mathbb{J}_3 +\frac{1}{2}\right]
\eeq
 where
\beq
[\lambda]\equiv \frac{\q^{\lambda/2}-\q^{-\lambda/2}}{\q^{1/2}-\q^{-1/2}}
\eeq
 The asymptotic massive particles form a spin-$s$ representation of $\uqsu2$.

This $S$-matrix can be expressed similarly as in \eqref{undef},
\bea
\S(\theta)=\sigma\left(\mathsf{P}\,\sum_{J=0}^{2s}f_{\q}^{[J]}(\theta)\,\mathbb{P}_{\q}^{[J]}\right)
\sigma^{-1},
\label{defS2}
\eea
but now  with some trigonometric scalar functions $f_{\q}^{[J]}(\theta)$,
$\q$-deformed projectors $\mathbb{P}_{\q}^{[J]}$ of $\uqsu2$,
and some gauge transformation $\sigma$.
All these ingredients of the $S$-matrices will be determined completely by imposing
constraints such as the Yang-Baxter equation, unitarity, and crossing symmetries.



For a generic $\q$, not a root of unity, the Lusztig-Rosso theorem states that 
the irreducible representations of the $\uqsu2$ are in one to one correspondence to those of $\mathfrak{su}_2$, and labelled by integer or half-integer $J$ \cite{Lusztig}.
The tensor products of two irreducible representations are decomposed into a direct sum of
other irreducible ones in the same way as the usual addition of two angular momenta in $\mathfrak{su}_2$.
The coefficients of this decomposition are now the quantum Clebsch-Gordan coefficients (qCGs), from which it is possible to construct the quantum projectors:
\bea
{\mathbb{P}_{\q}^{[J]}}_{m_1 m_2}^{m'_1 m'_2}=\sum_{M=-J}^{J}\langle s,m'_1;s,m'_2\vert J,M\rangle_{\q}
\langle J,M\vert s,m_1;s,m_2\rangle_{\q}.
\eea
Here, $\vert J,M\rangle$ is an eigenvector of the $\uqsu2$ Casimir operator  $\mathbb{Q}$ and of
$\mathbb{J}_3$
\beq
\mathbb{Q}|J,M\rangle = \left[J+\frac{1}{2}\right]|J,M\rangle \qquad,\qquad\mathbb{J}_3|J,M\rangle = M|J,M\rangle\,.
\eeq
We need explicit expressions of the qCGs to write down concrete $S$-matrices.
They are given by \cite{Kirillov,Hou,Ruegg}
\bea
&&\langle s,m_1;s,m_2\vert J,M\rangle_{\q}=f(J)\cdot \q^{(2s-J)(2s+J+1)/4+s(m_2-m_1)/2}
\label{qcg1}\\
&\times&\left\{[s+m_1]![s-m_1]![s+m_2]![s-m_2]![J+M]![J-M]!\right\}^{1/2}\,\sum_{\nu\ge 0}(-1)^{\nu}
\frac{\q^{-\nu (2s+J+1)/2}}{\mathcal{D}_{\nu}},\nonumber
\eea
where
\bea
\mathcal{D}_{\nu}&=&[\nu]![2s-J-\nu]![s-m_1-\nu]![s+m_2-\nu]![J-s+m_1+\nu]![J-s-m_2+\nu]!,
\nonumber\\
f(J)&=&\left\{\frac{[2J+1]([J]!)^2[2s-J]!}{[2s+J+1]!}\right\}^{1/2}.
\label{qcg2}
\eea
Here we use a convention of the $\q$-factorial for a positive integer $n$ 
\bea
[n]!=[n][n-1]\cdots[1],\quad [0]!=1,\quad [-n]!=\infty.
\eea
The summation over $\nu$ is bounded above since $D_{\nu}=\infty$ if any argument of $\q$-factorials
in $D_{\nu}$ is negative. 
From these expressions, one can compute the $\q$-projectors straightforwardly.

Notice that there may be problems in this qCG expression when $\q$ is a $n$-th root of unity:  $\q=\q(r,n)=e^{2\pi i r/n}$ with $n\in\mathbb{Z}_{>0}$ and $r=1,...,n-1$. Then for any integer $k$ multiple of $n$ the corresponding quantum number $[k]=0$. This fact would create diverging factors in the expressions (\ref{qcg1}) and (\ref{qcg2}). To avoid them, one has to resort to the more general formulae illustrated in \cite{Hou}, where suitable finite expressions for the qCG coefficients are recovered by a careful choice of normalization of the states and by a limiting procedure where a generic (non-root of unity) value of $\q$ approaches a root of unity value $\q\to\q(r,n)$. By this procedure, the root of unity cases leads to an expression for the quantum projectors related to that of generic $\q$ by continuity. So the formulae below can be taken as valid for any value of $\q$ on the unit circle, be it a root of unity or not.

%
%
From the Yang-Baxter equation, the scalar functions $f_{\q}^{[J]}(\theta)$ in \eqref{defS2} are 
obtained as
\bea
f_{\q}^{[J]}(\theta)=S_0(\theta)\,\left[\prod_{k=1}^{J}\frac
{\q^{k}-\q^{\theta/2\pi i}}{\q^{k}\q^{\theta/2\pi i}-1}\right],\quad J=0,1,\cdots,2s,
\eea
where an overall function $S_0(\theta)$ is still not fixed.

We define the charge conjugation $\mathsf{C}$ by
\bea
\mathsf{C}(A_m)=(-1)^{2s+m}A_{-m},\quad m=s,s-1,\cdots,-s
\eea
which can be represented by the following matrix 
\bea
\mathsf{C}=(-1)^s\left(
\begin{array}{ccccc}
0&0&\cdots&0&1\\
0&0&\cdots&-1&0\\
\vdots&\vdots&\ddots&\vdots&\vdots\\
0&(-1)^{2s-1}&\cdots&0&0\\
(-1)^{2s}&0&\cdots&0&0
\end{array}
\right),\qquad
\mathsf{C}^2=\mathbb{1}.
\eea
With this, the crossing symmetry is expressed as
\bea
\S^{{\rm t}_1}(\theta)=\mathsf{C}_1\cdot \S(i\pi-\theta)\cdot \mathsf{C}_1,\qquad
\mathsf{C}_1=\mathsf{C}\otimes\mathbb{1},
\label{crossing}
\eea
where ${\rm t}_1$ stands for the matrix transpose on the first vector space. 
To satisfy this relation, we need the gauge transformation $\sigma$
as noticed for the sine-Gordon model in \cite{BerLeC}.
In our case, it is given by
\bea
\sigma=\q^{\mathbb{J}_3\,\theta_1/2\pi i}\otimes\q^{\mathbb{J}_3\,\theta_2/2\pi i},
\eea
where $\theta_1$ and $\theta_2$ are defined in \eqref{defS}.
In addition to the crossing symmetry, one can show that this $S$-matrix is also invariant under
the charge conjugation $\mathsf{C}$, a parity $\mathsf{P}$, and a time reversal $\mathsf{T}$:
\bea
\S^{cd}_{ab}=\S^{{\bar c}{\bar d}}_{{\bar a}{\bar b}}=\S^{dc}_{ba}=\S^{ab}_{cd}.
\label{CPT}
\eea

Now we introduce a coupling constant $\gamma$ by
\bea
\q=e^{2\pi i\gamma}.
\eea
The scalar functions now can be expressed as
\bea
f_{\q}^{[J]}(\theta)=S_0(\theta)\,\prod_{k=1}^{J}\frac
{\sinh\left[\gamma(i k\pi-\theta)\right]}{\sinh\left[\gamma(i k\pi+\theta)\right]},\quad J=0,1,\cdots,2s.
\label{scalarf}
\eea

Next, we fix the overall scalar function $S_0(\theta)$ following a standard procedure.
By requiring unitarity and crossing symmetry, this function should satisfy
\bea
S_0(\theta)S_0(-\theta)=1,\qquad 
S_0(i\pi-\theta)=\prod_{k=1}^{2s}\frac{\sinh\left[\gamma(i (k+1)\pi-\theta)\right]}
{\sinh\left[\gamma(i k\pi+\theta)\right]}\,S_0(\theta).
\eea
The standard procedure for fixing $S_0$ is to express this as an infinite product of factors that satisfy 
the crossing symmetry and unitarity alternatingly as follows:
\bea
S_0(\theta)=-\prod_{k=1}^{2s}\left[\frac{\sinh\left[\gamma(i\pi k+\theta)\right]}
{\sinh\left[\gamma(i\pi k-\theta)\right]}\left(\prod_{\ell=1}^{\infty}
\frac{\sinh\left[\gamma(i\pi (k+\ell)-\theta)\right]\sinh\left[\gamma(i\pi (k-\ell)-\theta)\right]}
{\sinh\left[\gamma(i\pi (k+\ell)+\theta)\right]\sinh\left[\gamma(i\pi (k-\ell)+\theta)\right]}
\right)\right].\ \ 
\eea

When $s$ is an integer, i.e.  even $2s$, this infinite product is very much simplified  to
\bea
S_0(\theta)=-\prod_{m=1}^{s}
\frac{\sinh\left[\gamma(i2m\pi+\theta)\right]}
{\sinh\left[\gamma(i2m\pi-\theta)\right]}.
\eea
The $S$-matrix element $\S^{ss}_{ss}$ describing scattering between $A_s$ particles
with $\mathbb{J}_3=s$
can be read off from \eqref{scalarf} since only the 
projector $\mathbb{P}^{[2s]}$ contributes:
\bea
\S^{ss}_{ss}(\theta)=-\prod_{m=1}^{s}\frac{\sinh\left[\gamma(i(2m-1)\pi-\theta)\right]}
{\sinh\left[\gamma(i(2m-1)\pi+\theta)\right]}.
\label{even}
\eea
This reproduces the $S_{++}^{++}$ element of the sausage model for $s=1$.

For a half-integer $s$, \emph{i.e.} odd $2s$, one can convert the infinite products of trigonometric functions 
into products of $\Gamma$-functions
\bea
S_0(\theta)&=&-\prod_{m=1}^{2s}\Biggl\{\frac{1}{i\pi}\sinh\left[\gamma(\theta+im\pi)\right]
\Gamma\left[1-\gamma(m-1)+\frac{i\gamma\theta}{\pi}\right]
\Gamma\left[1-\gamma m-\frac{i\gamma\theta}{\pi}\right]\times\nonumber\\
&&\qquad\times\prod_{n=1}^{\infty}\left[
\frac{R_n^{[s,m]}(\theta)R_n^{[s,m]}(i\pi-\theta)}{
R_n^{[s,m]}(0)R_n^{[s,m]}(i\pi)}\right]\Biggr\},\\
R_n^{[s,m]}(\theta)&=&
\frac{\Gamma\left[\gamma(4sn-4s+2m-1)-\frac{i\gamma\theta}{\pi}\right]
\Gamma\left[1+\gamma(4sn-2m+1)-\frac{i\gamma\theta}{\pi}\right]}
{\Gamma\left[\gamma(4sn-2s+2m-1)-\frac{i\gamma\theta}{\pi}\right]
\Gamma\left[1+\gamma(4sn-2s-2m+1)-\frac{i\gamma\theta}{\pi}\right]}.
\eea
Again, the $S$-matrix element between $A_s$ particles becomes
\bea
\S^{ss}_{ss}(\theta)&=&-\prod_{m=1}^{2s}\Biggl\{\frac{1}{i\pi}\sinh\left[\gamma(\theta-im\pi)\right]
\Gamma\left[1-\gamma(m-1)+\frac{i\gamma\theta}{\pi}\right]
\Gamma\left[1-\gamma m-\frac{i\gamma\theta}{\pi}\right]\times\nonumber\\
&&\qquad\times\frac{\Gamma[\gamma m]}{\Gamma[1-\gamma(m-1)]}
\prod_{n=1}^{\infty}\left[\frac{R_n^{[s,m]}(\theta)R_n^{[s,m]}(i\pi-\theta)}{
R_n^{[s,m]}(0)R_n^{[s,m]}(i\pi)}\right]\Biggr\}.
\label{odd}
\eea

Although \eqref{even} for an integer $s$ and \eqref{odd} for a half-integer $s$ look very different, 
it turns out that both have the same integral representation for all $s$
\bea
\S^{ss}_{ss}(\theta)=-\exp\int_{-\infty}^{\infty}\frac{dk}{k}
\frac{\sinh(\pi ks)\sinh\pi k(s-\frac{1}{2\gamma})}{\sinh\frac{\pi k}{2\gamma}\sinh\pi k}
e^{ik\theta}.
\label{common}
\eea

From this representation, one can notice that 
\bea
\S^{ss}_{ss}(\theta)=-1,\qquad{\rm when}\quad \gamma=\frac{1}{2s}.
\label{freepoint}
\eea
This can be thought of as a kind of free point.
For $s=1/2$, this expression reduces to the prefactor of the sine-Gordon $S$-matrix in \cite{zamzam}.

In terms of this scalar factor, the $S$-matrix can be written as
\bea
\label{matS}
\mathsf{S}(\theta)=\S^{ss}_{ss}(\theta)\cdot \mathsf{S}_{\rm mat}(\theta),\quad
\mathsf{S}_{\rm mat}(\theta)\equiv \sigma\left(\mathsf{P}\,\sum_{J=0}^{2s}\,
\left[\prod_{k=J+1}^{2s}\frac
{\sinh\left[\gamma(i k\pi+\theta)\right]}{\sinh\left[\gamma(i k\pi-\theta)\right]}\right]
\mathbb{P}_{\q}^{[J]}\right)
\sigma^{-1}.
\eea

In the interval
\bea
0\leq \gamma\leq\frac{1}{2s}
\eea
the $S$-matrix does not present any pole in the physical strip $0\leq \mathrm{Im}\theta\leq\pi$ for any $s$, i.e. there are no bound states. We are in a repulsive regime. An analysis of these $S$-matrices for the attractive regime ($\gamma>1/2s$) should be very interesting, but it is out of the scope of the present paper.

We present explicit expressions for the next simplest $s=3/2$ $S$-matrix  which has 4 particles 
$A_m, m=3/2,1/2,-1/2,-3/2$ with $\mathsf{C}(A_m)=\overline{A}_m=A_{-m}$.
Denoting these particles with index $1,2,3,4$, hence $\bar{1}=4,\bar{2}=3$, 
non-vanishing $S$-matrix elements are given by
the prefactor in \eqref{common} multiplied by the following matrix elements:
\bea
\label{exps32}
&&\S_{11}^{11}=1,\ \S_{12}^{12}=\frac{(0)}{(3)},\ \S_{12}^{21}=\frac{s_{3}}{(3)},\
\S_{13}^{13}=\frac{(0)(-1)}{(2)(3)},\ \S_{13}^{22}=\frac{s_2\sqrt{s_3/s_1}(0)}{(2)(3)},\nonumber\\
&&\S_{13}^{31}=\frac{(s_1 s_4+2s_2)(0)}{(2)(3)},
\ \S_{22}^{22}=\frac{f_1}{(2)(3)},\ \S_{14}^{14}=\frac{(0)(-1)(-2)}{(1)(2)(3)},
\ \S_{14}^{23}=\frac{s_3(0)(-1)}{(1)(2)(3)},\nonumber\\
&&\ \S_{14}^{32}=\frac{s_2 s_3(0)}{(1)(2)(3)},
\ \S_{14}^{41}=\frac{s_1s_2s_3}{(1)(2)(3)},\ \S_{23}^{23}=\frac{(0)f_1}{(1)(2)(3)},\ 
\S_{23}^{32}=\frac{s_2f_2}{(1)(2)(3)},
\eea
and those related by $\mathsf{C},\mathsf{P},\mathsf{T}$ transformations given in \eqref{CPT}.
We have used the short notation
\bea
&&(n)\equiv 2\sinh\left[\gamma(\theta-i\pi n)\right],\quad s_n\equiv 2\sinh(in\pi\gamma),\ \nonumber\\
&&f_1=2\cosh\left[\gamma(2\theta-i\pi)\right]+\frac{s_{10}}{s_5}-2\frac{s_2}{s_1},\
f_2=2\frac{s_2}{s_1}\cosh\left[\gamma(2\theta-i\pi)\right]+s_2^2-2s_1^2-4.\nonumber
\eea



\setcounter{equation}{0}
\section{Thermodynamic Bethe ansatz}\label{sec3}
At a finite temperature, a large number of asymptotic particles can be created from the heat bath, 
carrying all possible momenta and $\mathbb{J}_3$ quantum numbers.
During elastic scattering processes, these particles will reach a thermal equilibrium where the
momenta are distributed in such a way that the free energy of the system is minimized. 
This condition for the equilibrium is the thermodynamic Bethe ansatz (TBA) equation.
The main technical difficulty arises from the fact that the $S$-matrix is nondiagonal.
Many different ``magnons'' can appear in diagonalizing transfer matrices.
To simplify the analysis, we consider, in this paper, $1/\gamma$ to be integers only, i.e. we are at values of $\q$ corresponding to primitive roots of unity. The more generic case should be approached by an adaptation of the Takahashi Suzuki decomposition methods \cite{TakSuz}, on which we intend to return in the future.

\def\N{{\mathcal{N}}}
\def\M{{\mathcal{M}}}
\subsection{Bethe-Yang equation}
If a number $\N$ of on-shell particles are created at a finite temperature $T$,
each of the momenta carried by these particles should satisfy a periodic boundary condition,
sometimes called the Bethe-Yang equation.
When the $S$-matrix  is nondiagonal,  this equation is given by a large tensor product of $\N$ $S$-matrices, called in \cite{alyosha_tba_rsos} ``color" transfer matrix $\mathbb{T}$,
formally equivalent to the ``inhomogeneous'' transfer matrix of an XXZ integrable spin chain 
with higher spins \cite{Kir-Resh1} 
\bea
e^{iL\mathsf{m}\sinh\theta_j}\mathbb{T}(\theta_j\vert\{\theta_i\})&=&1,\\
\mathbb{T}(\theta_j\vert\{\theta_i\})_{m_1,\cdots,m_\N}^{m'_1,\cdots,m'_\N}&=&
\sum_{n_1,\cdots,n_\N}\mathsf{S}_{n_1m_1}^{n_2m'_1}(\theta_1-\theta_j)
\mathsf{S}_{n_2m_2}^{n_3m'_2}(\theta_2-\theta_j)
\cdots\mathsf{S}_{n_{N}m_\N}^{n_1m'_\N}(\theta_\N-\theta_j).
\eea
Here $L$ is the volume of (infinite) one-dimensional space and we will eventually take the $L\to\infty$ limit.

As shown in \eqref{matS}, the color transfer matrix is factorized into a product of scalar functions $\S^{ss}_{ss}$ and a matrix part $\S_{\rm mat}$.
The matrix part has been diagonalized by analytic Bethe ansatz in \cite{Kir-Resh1,Kir-Resh2} for the spin $s$ XXZ chain and its generalization to the inhomogeneous case is straightforward.
The resulting Bethe-Yang equation for the eigenvalues of $\mathbb{T}$ is given by
\bea
e^{iL\mathsf{m}\sinh\theta_j}\prod_{k=1,k\neq j}^\N \S^{ss}_{ss}(\theta_j-\theta_k)\prod_{\ell=1}^\M
\mathsf{e}_{2s}(\theta_j-\lambda_{\ell})=1,
\label{BeYaE}
\eea
where we denote $\M$ as  the number of Bethe roots and use short notations
\bea
\mathsf{e}_{n}(\theta)\equiv\frac{\sinh\gamma(\theta+i\pi n/2)}{\sinh\gamma(\theta-i\pi n/2)},\quad
\mathsf{g}_{n}(\theta)\equiv\frac{\cosh\gamma(\theta+i\pi n/2)}{\cosh\gamma(\theta-i\pi n/2)}.
\eea
The parameters (Bethe roots) $\lambda_\ell$, often called ``magnonic rapidities'' in this context, must satisfy the Bethe ansatz equations (BAEs)
\bea
\label{BAEs}
\prod_{j=1}^\N\mathsf{e}_{2s}(\lambda_{\ell}-\theta_j)=
\prod_{k=1,k\neq\ell}^\M\mathsf{e}_{2}(\lambda_{\ell}-\lambda_k).
\eea

In the thermodynamic limit where we take $L\to\infty$ and $\N,\ \M\to\infty$, the Bethe roots organize into ``strings"
of length $n$, where the $n$ rapidities have the same real part but different imaginary values as 
\bea
\label{stringI} 
\lambda_{j,\alpha}^{(n)}=\lambda_{j}^{(n)}+\frac{i\pi}{2}(n+1-2\alpha),\quad \alpha=1,2,\cdots,n,
\eea
with the ``center'' of the string $\lambda_{j}^{(n)}$ to be real.
If the deformation parameter $\gamma$ is irrational, there is no limit on the length $n$, hence we need
to consider infinitely many different lengths of strings.
This makes the analysis of TBA equations very complicated.

For simplicity, we will restrict our considerations in this paper to
\bea
\gamma=\frac{1}{N},\qquad N\in\mathbb{Z},\quad  N\geq 2s+1,
\eea
which make the functions in \eqref{BAEs} periodic in the imaginary direction with period $\pi N/2$. Following \cite{TakSuz}, two types of strings are allowed, defined as follows:
\begin{itemize}
\item Type I: $\lambda_{j,\alpha}^{(n)}$ as in \eqref{stringI} with $n=1,2,\cdots,N-1$
\item Type II: $\lambda^{(N)}_{j}=\lambda_{j}+i\pi N/2$.
\end{itemize}
The $\M$ Bethe roots can be reorganized into a $\M_n$ number of type I strings of length $n=1,2,\cdots,N-1$ and a $\M_N$ number of type II strings.
The formation of strings rearranges the Bethe-Yang equation \eqref{BeYaE} as
\bea
e^{iL\mathsf{m}\sinh\theta_j}\prod_{k=1,k\neq j}^\N S_{00}(\theta_j-\theta_k)\,
\prod_{n=1}^{N-1}\left[\prod_{\ell=1}^{\M_n}S_{0n}(\theta_j-\lambda^{(n)}_{\ell})\right]\,
\prod_{k=1}^{\M_N}
S_{0N}(\theta_j-\lambda^{(N)}_{k})=1,
\label{BeYaE2}
\eea
where
\bea
S_{00}(\theta)&=&\S^{ss}_{ss}(\theta),\label{s00}\\
S_{0n}(\theta)&=&S_{n0}(\theta)=\prod_{\alpha=1}^{n}\mathsf{e}_{2s}(\theta-\frac{i\pi}{2}(n+1-2\alpha))
=\prod_{j=1}^{{\rm min}(n,2s)}\mathsf{e}_{|n-2s|+2j-1}(\theta),\label{s0n}\\
S_{0N}(\theta)&=&S_{N0}(\theta)=\mathsf{g}_{2s}(\theta).\label{s0N}
\eea

The BAEs in \eqref{BAEs} should be also rearranged in terms of the strings in a similar way 
\bea
\label{BAEs2}
\prod_{k=1}^\N S_{n0}(\lambda_j^{(n)}-\theta_k)\prod_{m=1}^{N-1}\,\left[
\prod_{i=1,i\neq j}^{\M_m}S_{nm}(\lambda_j^{(n)}-\lambda_i^{(m)})\right]\,
\prod_{\ell=1}^{\M_N}S_{nN}(\lambda_j^{(n)}-\lambda_{\ell}^{(N)})&=&1,\\
\prod_{k=1}^\N S_{N0}(\lambda_j^{(N)}-\theta_k)\prod_{n=1}^{N-1}\,\left[
\prod_{i=1}^{\M_m}S_{Nn}(\lambda_j^{(N)}-\lambda_i^{(n)})\right]\,
\prod_{\ell=1,\ell\neq j}^{\M_N}S_{NN}(\lambda_j^{(N)}-\lambda_{\ell}^{(N)})&=&1,
\label{BAEs3new}
\eea
where ($n,m=1,\cdots,N-1$)
\bea
S_{nm}(\theta)&=&S_{mn}(\theta)
=\left[\mathsf{e}_{|n-m|}(\theta)\mathsf{e}_{n+m}(\theta)\prod_{j=1}^{{\rm min}(n,m)-1}
\mathsf{e}_{|n-m|+2j-1}(\theta)^2\right]^{-1},\label{snm}\\
S_{nN}(\theta)&=&S_{Nn}(\theta)=\left[\mathsf{g}_{n-1}(\theta)\mathsf{g}_{n+1}(\theta)\right]^{-1},
\label{snN}\\
S_{NN}(\theta)&=&\mathsf{e}_{-2}(\theta).\label{sNN}
\eea

Eqs.(\ref{BeYaE2}) and \eqref{BAEs2} can be interpreted as diagonal Bethe-Yang equations in terms of the
original asymptotic massive particles and $N$ species of magnons, that do not transport any energy or momentum, but only internal degrees of freedom (colors) of the multiplets of asymptotic particles, whose
effective ``diagonalized'' $S$-matrices are given by (\ref{s00} -\ref{s0N}) and (\ref{snm}- \ref{sNN}).

\subsection{Derivation of thermodynamic Bethe ansatz equations}
We can simplify the Bethe Yang equations \eqref{BeYaE2}, \eqref{BAEs2} and \eqref{BAEs3new} by treating type I and II strings in a unified way as follows:
\bea
\label{BeYaE3}
e^{i\mathsf{m}L\sinh\theta_j}\prod_{k=1,k\neq j}^\N S_{00}(\theta_j-\theta_k)\,
\prod_{n=1}^{N}\left[\prod_{\ell=1}^{\M_n}S_{0n}(\theta_j-\lambda^{(n)}_{\ell})\right]\,&=&1,\\
\label{BAEs3}
\prod_{k=1}^\N S_{n0}(\lambda_j^{(n)}-\theta_k)\prod_{m=1}^{N}\,\left[
\prod_{i=1,i\neq j}^{\M_m}S_{nm}(\lambda_j^{(n)}-\lambda_i^{(m)})\right]&=&1,\quad
n=1,\cdots,N.
\eea
In $L\to\infty$ limit, the densities of the particles and magnons are defined by
\beq
\label{densities}
\sigma_n(\theta)=\frac{1}{L}\frac{d\mathsf{n}_n}{d\theta},\qquad n=0,1,\cdots,N,
\eeq
where $d\mathsf{n}_n$ is the number of the massive particles ($n=0$) or the magnons of type I and II
which carry the rapidities between $\theta$ and $\theta+d\theta$.

In terms of these densities, we can rewrite the Bethe-Yang equations by taking logarithms on both sides,
\bea
\label{BetheYang}
\sigma_{n}(\theta)+{\tilde\sigma}_n(\theta)=\delta_{n0}\mathsf{m}\cosh\theta-\nu_n\sum_{m=0}^{N}
K_{nm}\star\sigma_{n}(\theta),\qquad n=0,1,\cdots,N,
\eea
where we used a short notation 
\bea
\nu_n=\begin{cases}
1,&n=0,N\\
-1,&n=1,\cdots,N-1
\end{cases}
\eea
and a standard convolution notation ($\star$) 
\bea
f\star g(\theta)=\int_{-\infty}^{\infty}f(\theta')g(\theta-\theta')d\theta',
\eea
along with the kernels defined by
\bea
K_{nm}(\theta)=\frac{1}{2\pi i}\frac{d}{d\theta}\ln S_{nm}(\theta).
\eea
The densities of ``holes" ${\tilde\sigma}_n$ are defined similarly as \eqref{densities} for ``unoccupied'' states.

The system of equations \eqref{BetheYang} can be simplified further by taking into consideration certain identities on the Fourier transforms of the kernels. The explicit expressions of these kernels and the technical details of the derivation of the relations among them can be found in \cref{appA}. The result is a set of simplified Bethe-Yang equations, expressed by a single ``universal" kernel with a simple structure of couplings among densities.

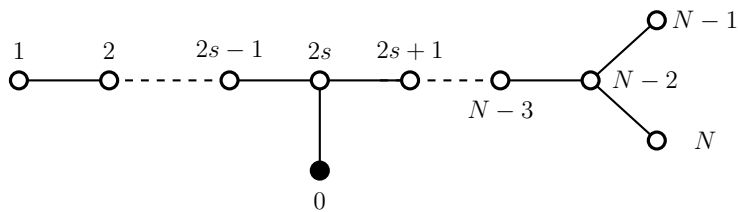
\begin{figure}[h]
    \centering
\begin{tikzpicture}[scale=.8, transform shape]
\draw[thick] (1.5,-2) -- (3,-2);
\draw[thick,dashed] (3,-2) -- (5,-2);
\draw[thick] (5,-2) -- (6.5,-2);
\draw[thick] (6.5,-2) -- (6.5,-3.5);
\draw[thick] (6.5,-2) -- (8,-2);
\draw[thick,dashed] (7.5,-2) -- (9.5,-2);
\draw[thick] (9.5,-2) -- (11,-2);
\draw[thick] (11,-2) -- (12.1,-1);
\draw[thick] (11,-2) -- (12.1,-3);

\filldraw[color=black, fill=white, very thick] (1.5,-2) circle (4pt);
\node [label] at (1.5,-1.5) {$1$};
\filldraw[color=black, fill=white, very thick] (3,-2) circle (4pt);
\node [label] at (3,-1.5) {$2$};
\filldraw[color=black, fill=white, very thick] (5,-2) circle (4pt);
\node [label] at (5,-1.5) {$2s-1$};
\filldraw[color=black, fill=black, very thick] (6.5,-3.5) circle (4pt);
\node [label] at (6.5,-4) {$0$};
\filldraw[color=black, fill=white, very thick] (6.5,-2) circle (4pt);
\node [label] at (6.5,-1.5) {$2s$};
\filldraw[color=black, fill=white, very thick] (8,-2) circle (4pt);
\node [label] at (8,-1.5) {$2s+1$};

\filldraw[color=black, fill=white, very thick] (9.5,-2) circle (4pt);
\node [label] at (9.5,-2.5) {$N-3$};
\filldraw[color=black, fill=white, very thick] (11,-2) circle (4pt);
\node [label] at (11.9,-2) {$N-2$};
\filldraw[color=black, fill=white, very thick] (12.1,-1) circle (4pt);
\node [label] at (12.9,-1) {$N-1$};
\filldraw[color=black, fill=white, very thick] (12.1,-3) circle (4pt);
\node [label] at (12.9,-3) {$N$};
\end{tikzpicture}
\caption{Dynkin-like structure of the TBA equations for $2s<N-1$. Note that the graph is a proper $D_{N+1}$ Dynkin diagram only for $2s=1$ and an extended one $\hat{D}_{N+1}$ for $2s=2 $.}
\label{graph1}
\end{figure}

\begin{figure}[ht]
    \centering
\begin{tikzpicture}[scale=.8, transform shape]
\draw[thick] (1.5,-2) -- (3,-2);
\draw[thick,dashed] (3,-2) -- (5,-2);
\draw[thick] (5,-2) -- (6.5,-2);
\draw[thick] (6.5,-2) -- (7.6,-1);
\draw[thick] (6.5,-2) -- (7.6,-3);
\draw[thick] (7.6,-1) -- (8.7,-2);
\draw[thick] (7.6,-3) -- (8.7,-2);

\filldraw[color=black, fill=white, very thick] (1.5,-2) circle (4pt);
\node [label] at (1.5,-1.5) {$1$};
\filldraw[color=black, fill=white, very thick] (3,-2) circle (4pt);
\node [label] at (3,-1.5) {$2$};
\filldraw[color=black, fill=white, very thick] (5,-2) circle (4pt);
\node [label] at (5,-2.6) {$N-3$};
\filldraw[color=black, fill=black, very thick] (8.7,-2) circle (4pt);
\node [label] at (9.1,-2) {$0$};
\filldraw[color=black, fill=white, very thick] (6.5,-2) circle (4pt);
\node [label] at (6.3,-1.5) {$N-2$};
\filldraw[color=black, fill=white, very thick] (7.6,-1) circle (4pt);
\node [label] at (7.6,-0.6) {$N-1$};
\filldraw[color=black, fill=white, very thick] (7.6,-3) circle (4pt);
\node [label] at (7.6,-3.4) {$N$};
\end{tikzpicture}
\caption{Structure of the TBA equations of $2s=N-1$.}
\label{graph2}
\end{figure}
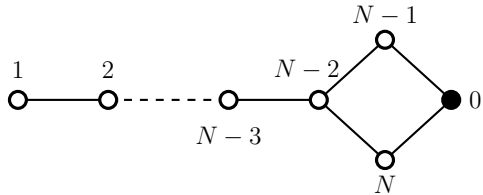

The TBA equations can be derived  in a standard procedure by minimizing the 
free energy for a finite temerature $T=1/R$ with the universal Bethe-Yang equations (\ref{sigmaeq2})--(\ref{sigmaeq6}) as constraints.
We will use the Fermi-Dirac statistics in our derivation since $\S_{ss}^{ss}(0)=-1$.
They are given by coupled nonlinear integral equations for pseudo-energies $\epsilon_n$
\begin{equation}
    \label{eq:TBAeq}
\epsilon_n(\theta)=\delta_{n,0}\mathsf{m}R\cosh\theta-\sum_{m=0}^{N}\mathbb{I}_{nm}\,p\star
\log\left(1+e^{-\epsilon_m}\right)(\theta),\qquad n=0,1,\cdots,N,
\end{equation}
where we have introduced the pseudo-energies
\begin{equation}
    \epsilon_0(\theta) = \log\frac{\tilde{\sigma}_0}{\sigma_0}, \qquad \epsilon_n(\theta) = \log\frac{\sigma_n}{\tilde{\sigma}_n}, \quad n=1,\dots,N-1, \qquad \epsilon_N(\theta) = \log\frac{\tilde{\sigma}_N}{\sigma_N},
\end{equation}
and the universal kernel
\begin{equation}
\label{unikern}
p(\theta)=\frac{1}{2\pi \cosh\theta}.
\end{equation}
Here $\mathbb{I}_{nm}$ are the matrix elements of the incidence matrix\footnote{It is the matrix whose element $m,n$ is 1 when the nodes $n$ and $m$ are connected, 0 otherwise.} of the graphs in \cref{graph1,graph2} when $2s<N-1$ and $2s=N-1$, respectively.

At finite temperature $T$, the free energy per unit length is obtained by the pseudo-energy $\epsilon_0$ using
\begin{equation}\label{freen}
\frac{f(T)}{T}=-\int_{-\infty}^{\infty}\frac{\mathsf{m}}{2\pi}\cosh\theta\ln\left(1+e^{-\epsilon_0(\theta)}\right)d\theta.
\end{equation}

We want to emphasize that this universal TBA is possible thanks to a remarkable relation \eqref{KR00} between the minimal scalar factor $S_0$ and other scattering amplitudes of the magnons. 
If a CDD factor is added, this relation is not valid, therefore TBA cannot be written in the universal way.

\setcounter{equation}{0}
\section{Numerical Analysis}\label{sec4}
The TBA is based on the idea that there are two equivalent ways to quantize the theory along different channels. This allows us to identify the free energy per unit length of \eqref{freen} with the Casimir energy of the mirror theory,
\begin{equation}
    E_0(T) = \frac{f(T)}{T}.
\end{equation}
It is customary to parameterize the vacuum energy as $E_0(T) = - \widetilde{c}(r) T\pi/6 $, by introducing the so-called scaling function $\widetilde{c}(r)$, with $r=\textsf{m}R$ being a dimensionless parameter. 
From \cref{freen} one finds
\begin{equation}\label{ctilde}
    \widetilde{c}(r) = \frac{3}{\pi^2} \int_{-\infty}^{\infty} r \cosh(\theta) L_0(\theta) d\theta
\end{equation}
where $L_0(\theta)= \log(1+e^{-\epsilon_0(\theta)})$.
In the limit $r \to 0$, the ultraviolet (UV) limit, this function encodes all the relevant data of the underlying conformal field theory, since
\begin{equation}
    \lim_{r\to 0} \widetilde{c}(r) = c-24\Delta_{\min},
\end{equation}
where $c$ is the central charge and $\Delta_{\min}$ is the lowest eigenvalue of the zero-th Virasoro generator.
 
\begin{figure}
  \begin{minipage}{0.5\textwidth}
    \centering
    \includegraphics[width=\linewidth]{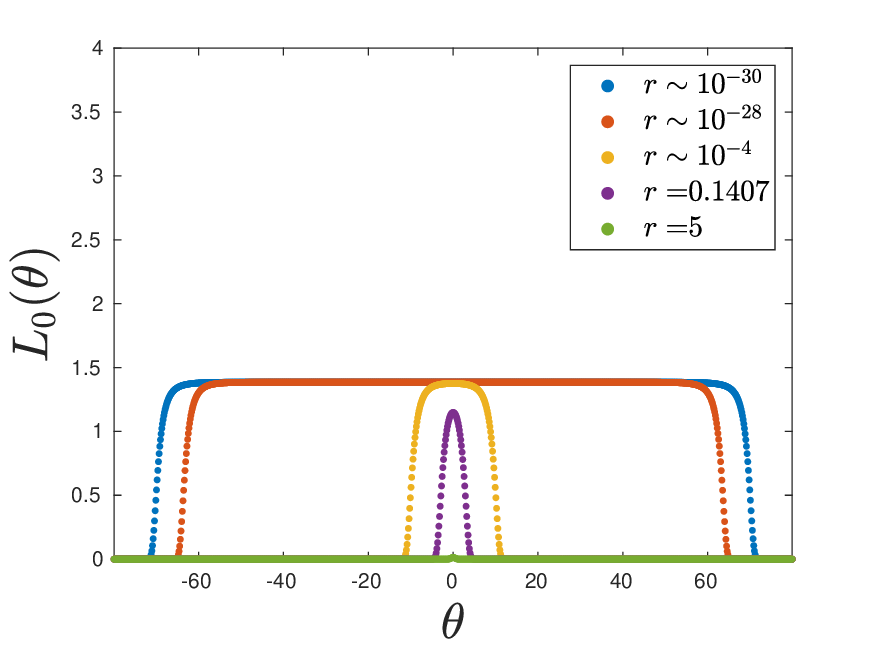}
  \end{minipage}
  \hfill
  \begin{minipage}{0.5\textwidth}
    \centering
    \includegraphics[width=\linewidth]{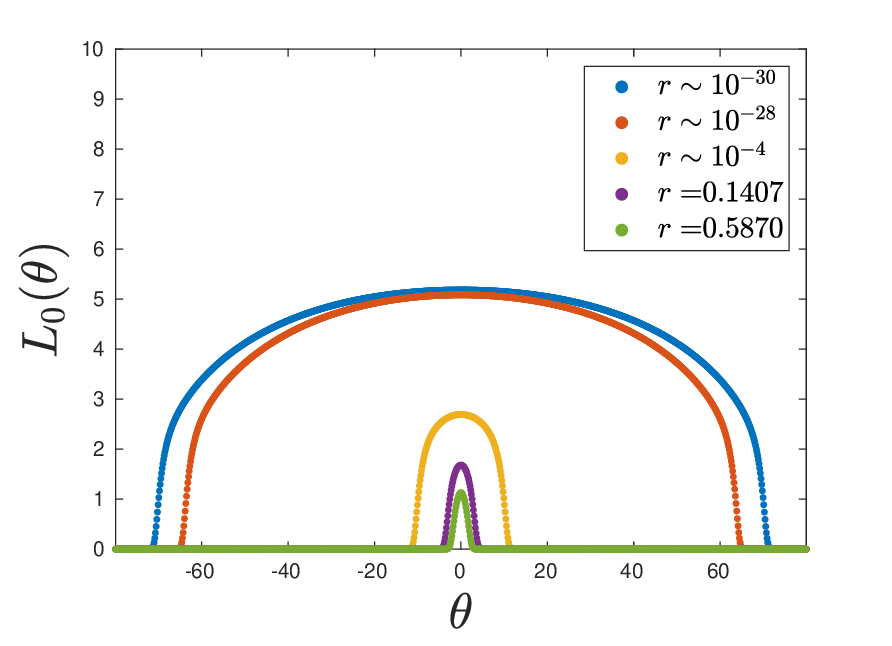}
  \end{minipage}
\caption{The functions $L_0(\theta)$ for spin $s=1/2$ (left) and $s=1$ (right) with $\gamma=1/7$, for different values of $r$. One can see that for smaller values of $r$ the plateau starts to form.}
\label{fig1}
\end{figure}

The TBA equations \eqref{eq:TBAeq} are a system of non-linear integral equations for which is in general very difficult to find an analytical closed solution.
Sometimes, however, it is possible to do so. For example, in the UV limit $r\to 0$, it is well known that for some theories it is possible to express the TBA in terms of simple algebraic equations whose explicit solutions can be used to derive the central charge $c$ in terms of sums of Rogers dilogarithms.\footnote{CDD factors may change these equations in such a way that no more real solutions are allowed.}
The fundamental property shared by these theories is the fact that as $r$ approaches 0, the functions $\log(1+e^{-\epsilon(\theta)})$, develop a plateau of width $\sim 2\log(2/r)$, as first noticed by Al. Zamolodchikov \cite{alyosha_tba}.
In the family of scattering theories we have introduced above, we have two well-known examples of this behavior: the sine-Gordon model, corresponding to spin $s=1/2$, and the sausage model for $s=1$. In these cases, the plateaus start to form for small values of $r$, see e.g. \cref{fig1}, and therefore one can explicitly compute the value of the central charge using dilogarithms obtaining $c=1$ and $c=2$, respectively, independently from the value of $\gamma=1/N$.

\begin{figure}[h]
  \begin{minipage}{0.5\textwidth}
    \centering
    \includegraphics[width=\linewidth]{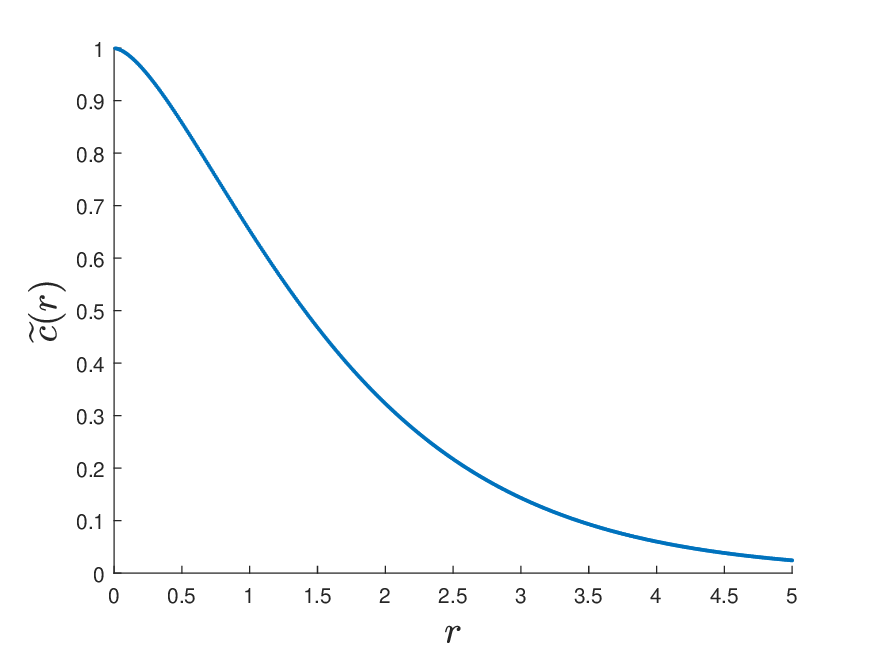}
  \end{minipage}
  \hfill
  \begin{minipage}{0.5\textwidth}
    \centering
    \includegraphics[width=\linewidth]{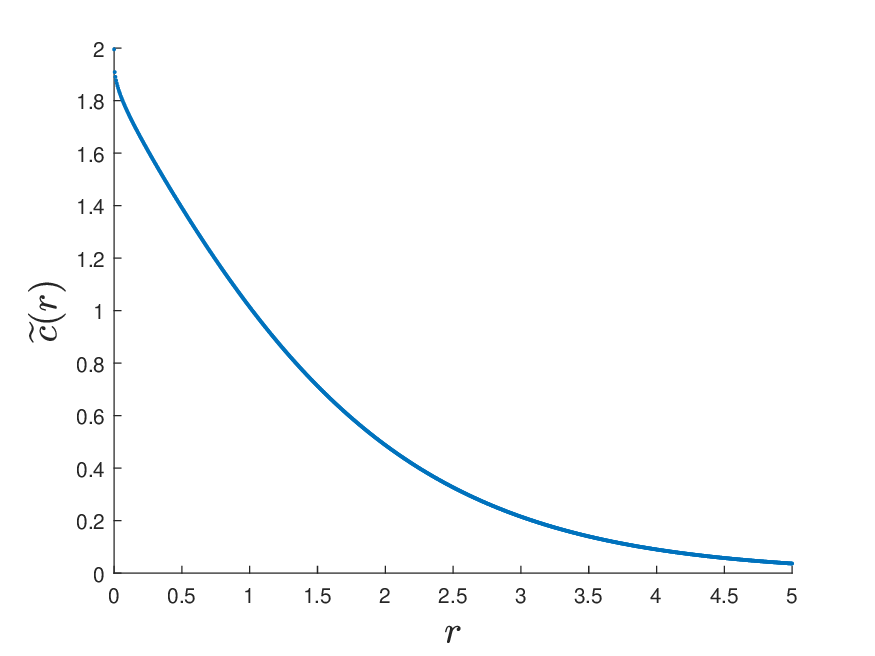}
  \end{minipage}
\caption{Scaling functions for spin $s=1/2$ (left) and $s=1$ (right).}
\label{fig2}
\end{figure}

As pointed out above, it is usually difficult to find a closed solution for generic $r$: for this reason, it becomes very useful to perform numerical analysis to study the behavior of these theories. The method that has proven to be more effective is by solving the system of equations via successive iterations. The idea is to start from the initial guess $\epsilon^{(0)}_n= (r\cosh\theta,0,\dots,0)$ for $n=0,\dots,N$ and then define the $k$-th iterative solution, with $k\geq 0$, as
\begin{equation}
    \epsilon_n^{(k+1)}(\theta) = \delta_{n,0} r \cosh(\theta) - \sum_{m=0}^N \mathbb{I}_{nm} (p\ast L^{(k)}_{m})(\theta), \qquad n=0,1,\dots,N,
\end{equation}
where $L^{(k)}_n(\theta)=\log(1+e^{-\epsilon^{(k)}_m(\theta)})$. In general, this process is not guaranteed to converge, but if it does, one is then able to find with arbitrarily high accuracy the values of the pseudo-energies and the corresponding $L_n(\theta)$ (an extensive study of this convergence problem has been done in \cite{ingo17}).

This allows us to compute numerically the integral \eqref{ctilde} at different values of $r$, finding the value of the scaling function and, possibly, of the central charge of the underlying conformal theory. 
The cases of $s=1/2$ and $s=1$ are shown in \cref{fig2}.

\subsection{Higher spins and Hagedorn-like singularity}
\begin{figure}[h!]
  \begin{minipage}{0.5\textwidth}
    \centering
    \includegraphics[width=\linewidth]{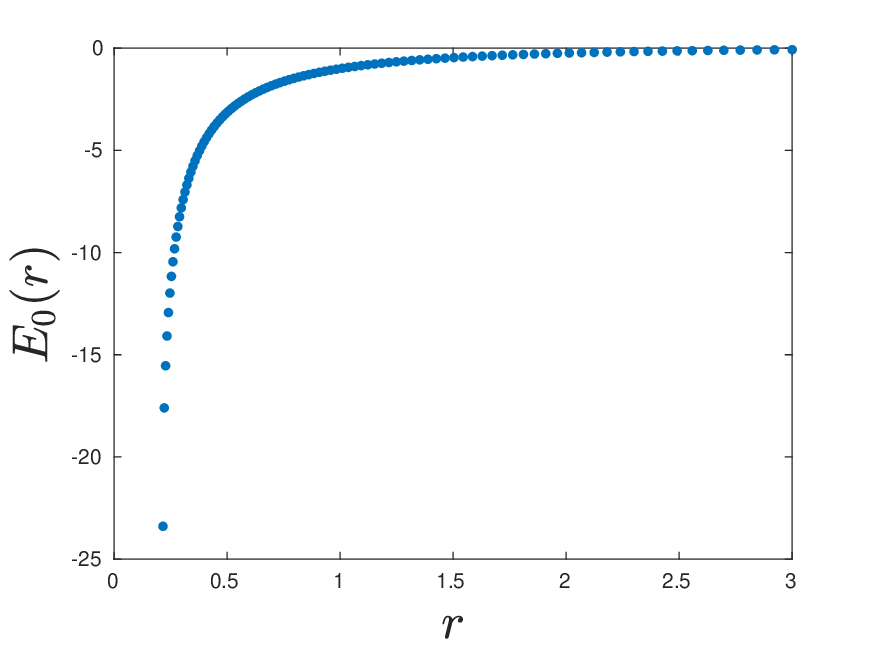}
  \end{minipage}
  \hfill
  \begin{minipage}{0.52\textwidth}
    \centering
    \includegraphics[width=\linewidth]{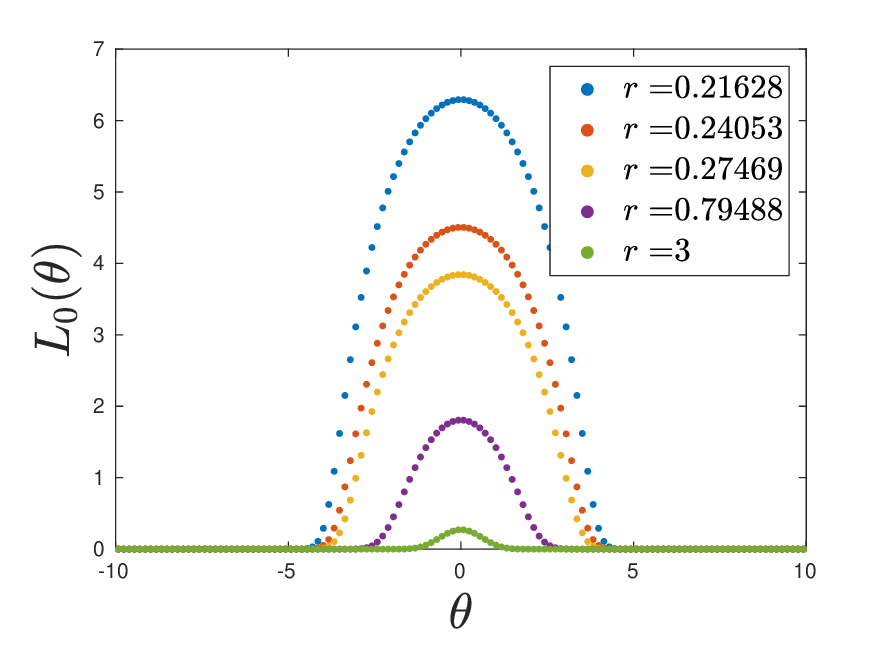}
  \end{minipage}
\caption{Left: the vacuum energy $E_0(r)$ as it approaches the singular point $r^*=0.21628(2)$; right: the kernel $L_0(\theta)$ at different values of $r$. Both were obtained for $s=5/2$ and $N=12$. }
\label{fig3}
\end{figure}

Having a natural generalization of the $S$-matrix for higher values of the spin, $s\geq \frac32$, and of the corresponding TBA equations, it is natural to ask what kind of theories they describe.
Performing the same iterative procedure as above, we observe an unexpected behavior as the ground state energy $E_0(r)$ diverges at a positive finite value $r^*$ and, correspondingly, that the functions $L_n(\theta)$ do not develop a plateau, but rather become more peaked around $\theta=0$ as they approach the singular value, as shown in \cref{fig3}.

Extending the numerical analysis to different values of the spin and of the coupling constant, we see that the critical value $r^*$ is a function of both $s$ and $N=1/\gamma$. Some values of $r^*$ are listed in \cref{table}.
\begin{table}[h]
\renewcommand{\arraystretch}{1.1} 
    \centering
    \begin{tabular}{c ||c|c|c|c}
         & $s=3/2$ & $s=2$ & $s=5/2$ & $s=3$\\
         \hline
      $ N=4 $ & 0.06024(4) & -         & -         & -\\
       $N=5$  & 0.01683(2) & 0.22505(9) & -         & -\\
      $ N=6 $ & 0.00722(5) & 0.09996(5) & 0.40380(3) & -\\
       $N=7$  & 0.00392(8) & 0.05976(6) & 0.21628(2) & 0.57301(7)\\
       $N=8$  & 0.00248(7) & 0.04195(5) & 0.14665(8) & 0.34110(6)\\
       $N=9$  & 0.00174(9) & 0.03255(2) & 0.11269(7) & 0.24773(3)\\
      $ N=10$ &  0.00132(7)& 0.02699(9) & 0.09349(6) & 0.19958(2)\\
       $N=11$  & 0.00106(6) & 0.0234(5) & 0.08157(4) & 0.17123(0)\\
       $N=12$  & 0.00089(4) & 0.02106(7) & 0.07367(8) & 0.15307(2)\\
    \end{tabular}
    \caption{Some values of the critical scale $r^*$ for different values of  $s$ and $\gamma=1/N$.}
    \label{table}
\end{table}
\begin{figure}[h!]
    \centering
    \includegraphics[width=\textwidth]{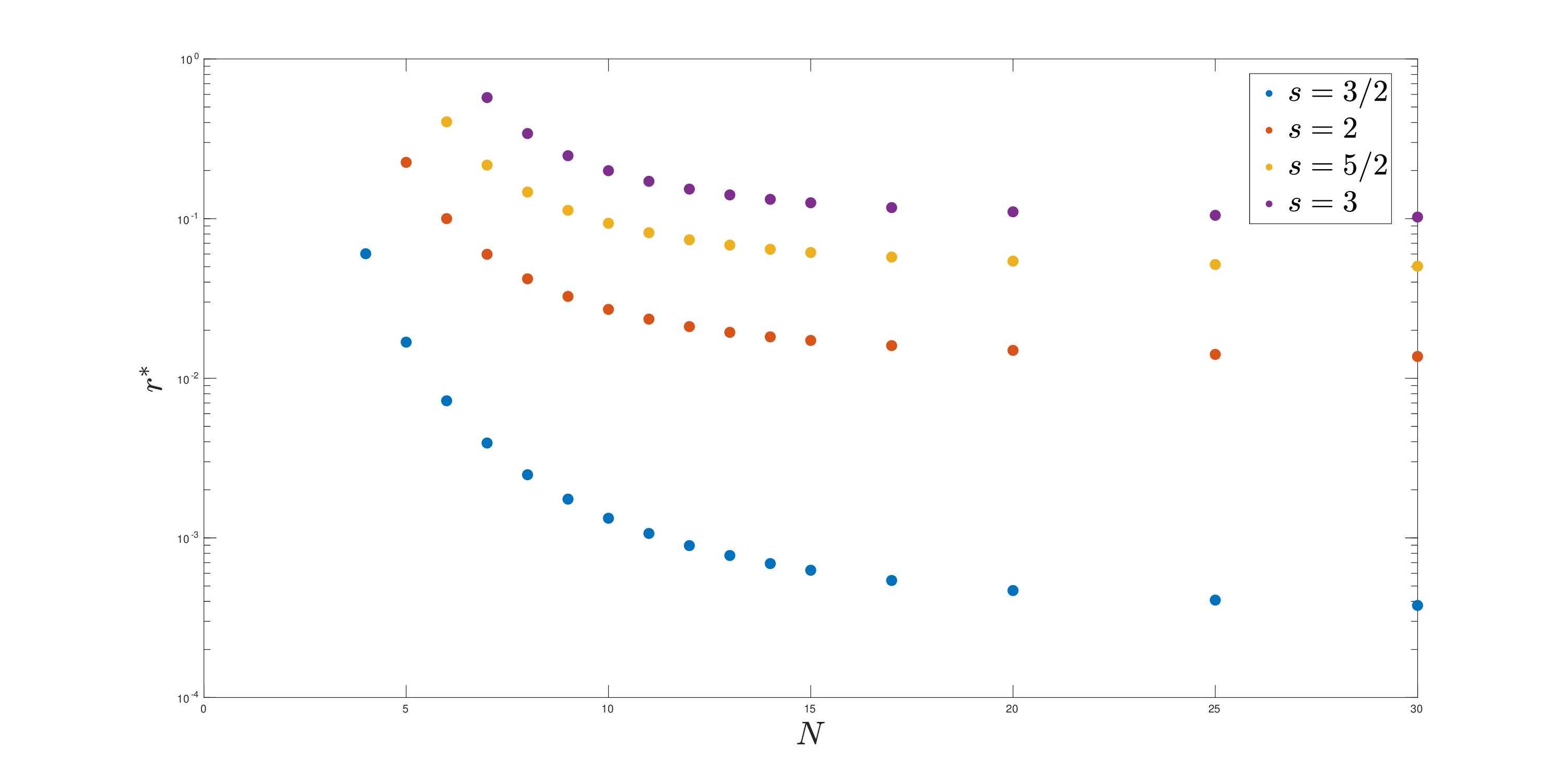}
    \caption{Value of the singular point $r^*$, for different values of spin and coupling constant $\gamma=1/N$. The values are computed with precision to the 6th decimal digit. The $r^*$-axis is log-scaled.}
    \label{fig4}
\end{figure}

Moreover, as can be seen from \cref{fig4}, the critical values $r^*$  seem to converge to non-zero values even for vanishing coupling constant
$\gamma=1/N\to 0$ for $s\ge 3/2$.
This means that the singularity does occur also at the $\mathfrak{su}(2)$ symmetric points and the UV limit does not exist even in those cases.\footnote{In \cite{AladimMartins} this computation has been performed by taking first the UV limit for a finite $N$ and then the $N\to\infty$ limit. Since there is no UV limit for finite $N$, this limiting procedure is inconsistent.}
It is interesting to notice that a Hagedorn transition of $\mathcal{N}=4$ SYM at finite temperature exists in the limit where the coupling constant vanishes, as shown by \cite{sundb}.

A similar behavior has been recently studied in $T\overline{T}$-deformed theories with detailed numerical analysis \cite{camilo}. In this case, the vacuum energy develops a square root singularity
\begin{equation}
    E_0(r) \sim_{r\to r^*} c_0 + c_{1/2}\sqrt{r-r^*}.
\end{equation}

In this setting, the singularity ultimately appears as a consequence of the presence of a CDD factor and it has been regarded as the appearance of a Hagedorn-like phase transition. Remarkably, it has been shown that by finely tuning the parameters of the deformation, one can ultimately remove the singularity, as described in \cite{AhnLeC}.

\begin{figure}[h!]
  \begin{minipage}{0.5\textwidth}
    \centering
    \includegraphics[width=\linewidth]{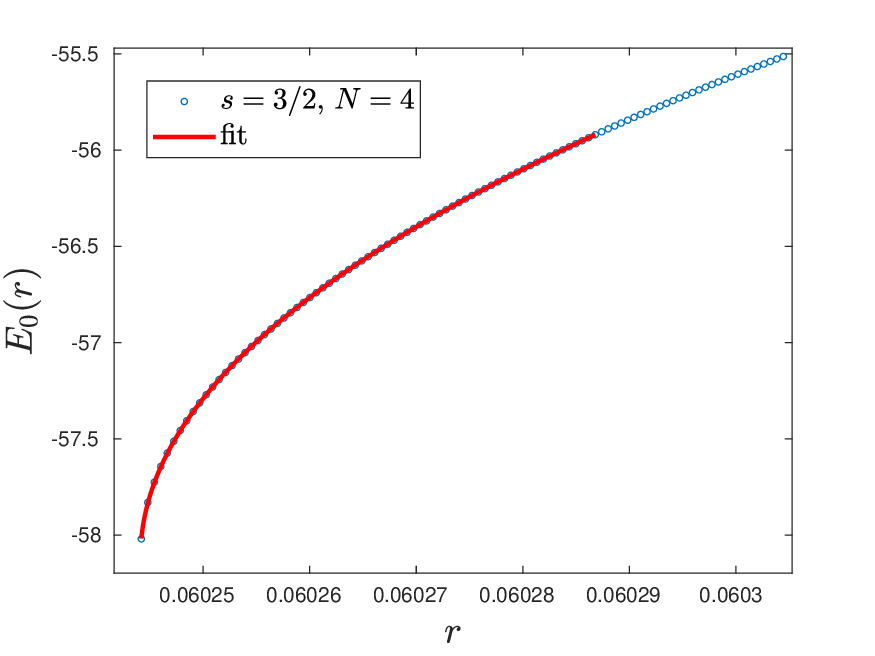}
  \end{minipage}
  \hfill
  \begin{minipage}{0.5\textwidth}
    \centering
    \includegraphics[width=\linewidth]{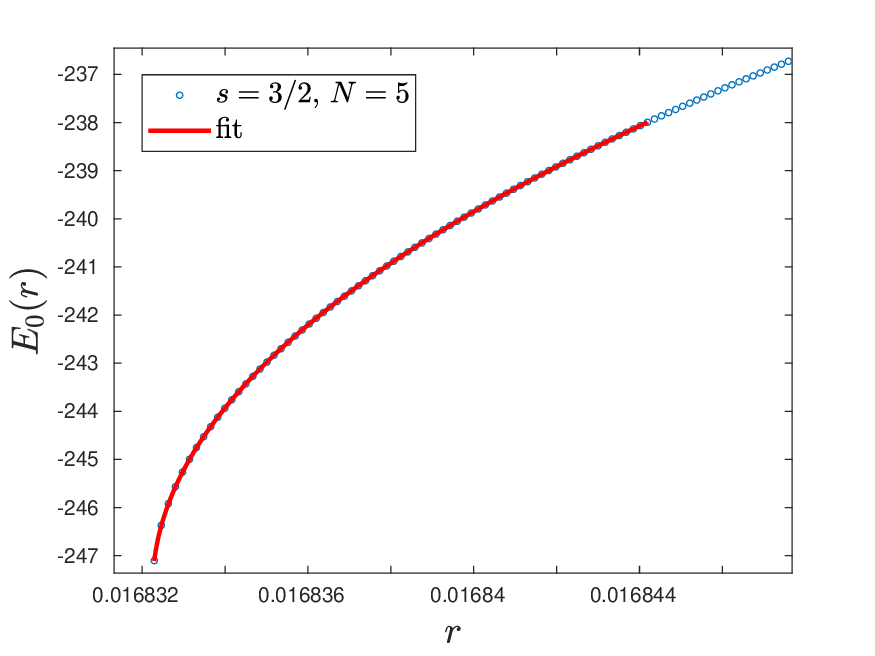}
  \end{minipage}
\caption{Examples of fitting for $s=3/2$ and $N=4$ (left) and $N=5$ (right).}
\label{fig5}
\end{figure}

The theories we have introduced in this work present some similar aspects, but they are crucially different. Indeed, the $S$-matrices we consider are not obtained as a deformation of some known theory but are genuinely obtained by imposing the defining properties of a scattering theory in two dimensions, as explained in \cref{sec2}. As a result, this singularity is in a sense more ``fundamental", as it cannot be removed by a fine-tuning of the parameters.
We analyzed the behavior of these models close to the singularity, for different values of the spin, at different values of the coupling constant. More explicitly, we have generated several points in a close neighborhood of width $\sim 1\% $ of the singular points of \cref{table}. We then used these data and fitted the curves, as shown in \cref{fig5} with a fitting function given by
\begin{equation}
   E_0^{\texttt{fit}}(r) =  b (r-r^*)^{a} +c_0.
\end{equation}

\begin{figure}[h!]
        \centering
        \begin{subfigure}[b]{0.49\textwidth}
            \centering
            \includegraphics[width=\textwidth]{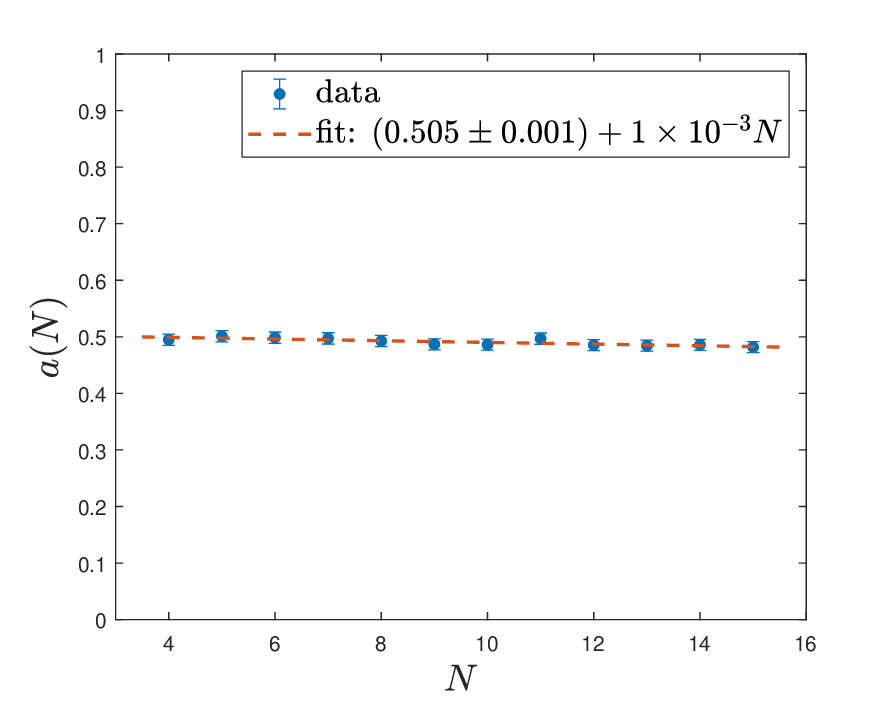}
            \caption[]%
            {{\small $s=3/2$}}
            \label{fit1}
        \end{subfigure}
        \hfill
        \begin{subfigure}[b]{0.49\textwidth}
            \centering
            \includegraphics[width=\textwidth]{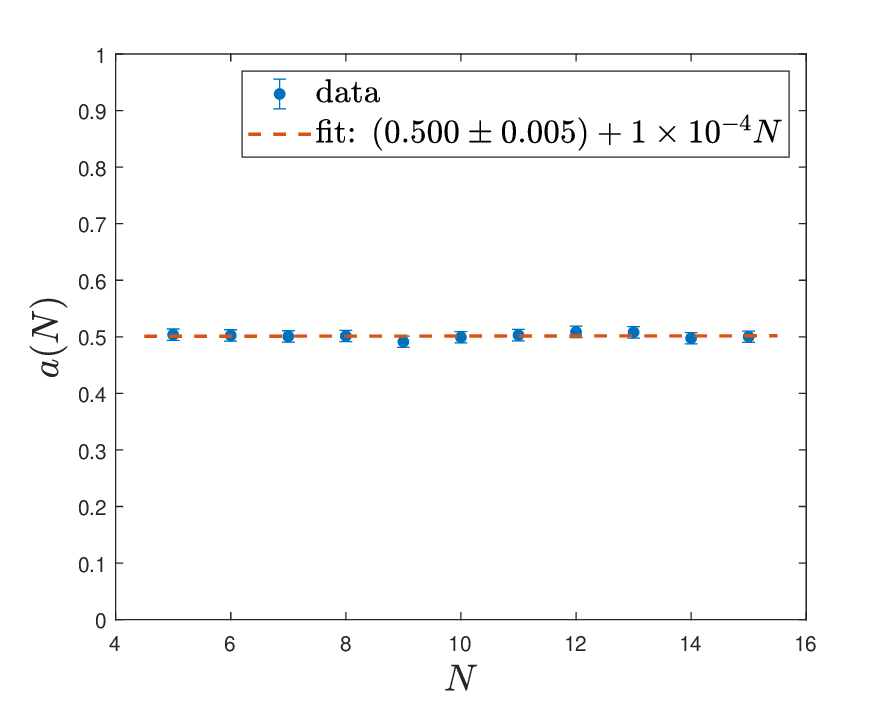}
            \caption[]%
            {{\small $s=2$}}
            \label{fit2}
        \end{subfigure}
        \vskip\baselineskip
        \begin{subfigure}[b]{0.49\textwidth}
            \centering
            \includegraphics[width=\textwidth]{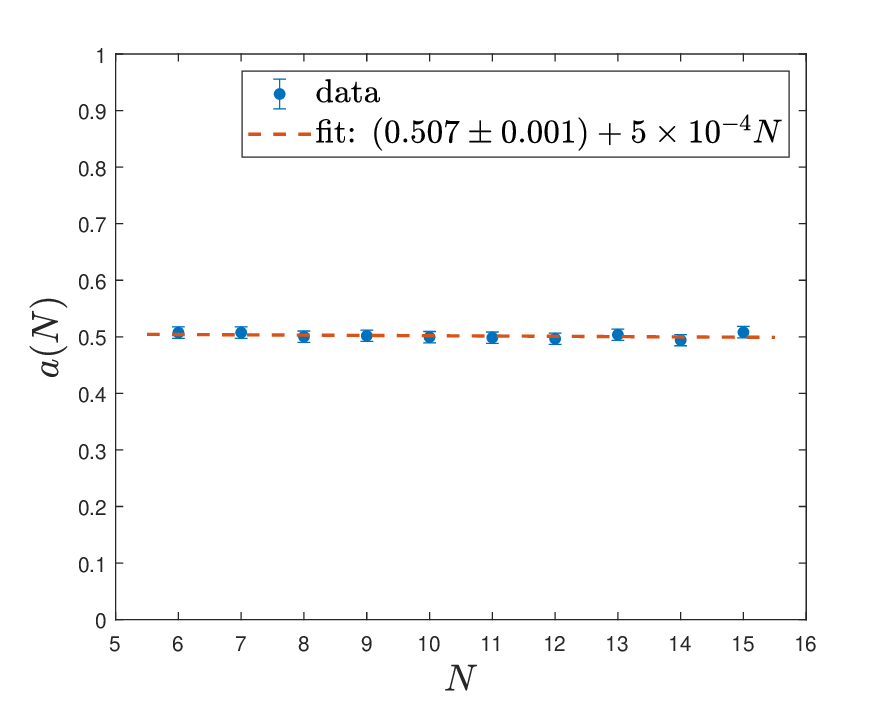}
            \caption[]%
            {{\small $s=5/2$}}
            \label{fit3}
        \end{subfigure}
        \hfill
        \begin{subfigure}[b]{0.49\textwidth}
            \centering
            \includegraphics[width=\textwidth]{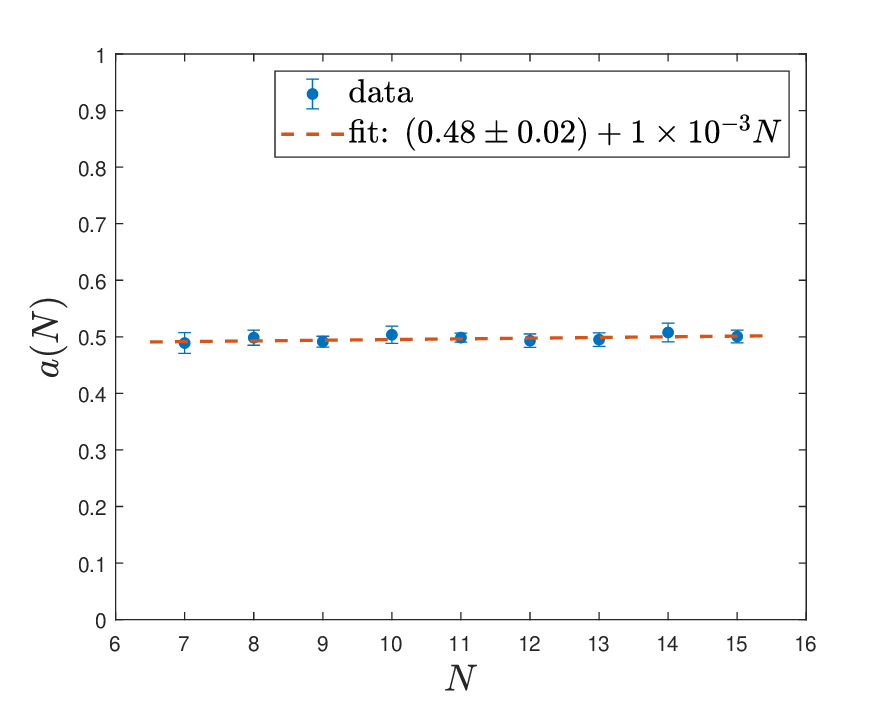}
            \caption[]%
            {{\small $s=3$}}
            \label{fit4}
        \end{subfigure}
        \caption{Fitted value of the exponent $a$, for different values of the spin.}
        \label{fig6}
    \end{figure}

In \cref{atable} we present the results of the critical exponent $a$ obtained from the numerical analysis, and in \cref{fig6} we show a fit of these points.
The main source of error in the exponent $a$ is from a sensitive dependence on the initial guesses used in the fitting algorithm. We performed the fit for different initial guesses to estimate the average value of $a$ and the associated error which is around $2\%$.

\begin{table}[h]
  \setlength\tabcolsep{4pt}
\renewcommand{\arraystretch}{1.1} 
    \centering
    \begin{tabular}[h]{c   c  c  c c c c c c c c c c c  }
      &\multicolumn{12}{c}{$N$} \\ \cmidrule{2-13}
       & 4 & 5 & 6 & 7 & 8 & 9 & 10 & 11 & 12 & 13 & 14 & 15 \\
      \hline
      $s=3/2$ &  0.495  &  0.501 &   0.498  &  0.497  &  0.492 &   0.487  &  0.486  &  0.497 &   0.486  &  0.485  &  0.488  & 0.482  \\
      \hline
      $s=2$ & |  &  0.504 &   0.502 &   0.501 &   0.501 &   0.491  &  0.499 &   0.504 &   0.509 &   0.508 &   0.497  &  0.500 \\
      \hline
      $s=5/2$ & | & | & 0.507 &   0.507  &  0.500 &   0.502 &   0.499 &   0.499 &   0.497 &   0.493 &   0.494  &  0.508\\
      \hline
      $s=3$ & | & | & | &  0.489 &    0.498 &   0.491 &   0.504 &   0.499 &   0.493 &   0.495&   0.508  &  0.501\\
      \bottomrule
    \end{tabular}
    \caption{Values of the fitted exponent $a$, for different values of the spin and $N=4,5,\dots,15$.
    }
    \label{atable}
\end{table}

Remarkably, we observe that it becomes independent of the value of the coupling constant and the spin, approaching a universal value of $\sim 1/2$, compatible with a square root behavior.

Similarly, in \cref{tables3} we present the fitted values of the parameters $b$ and $c_0$, only for the case $s=3$.
\begin{table}[h!]
  \setlength\tabcolsep{4pt}
\renewcommand{\arraystretch}{1.1} 
    \centering
    \begin{tabular}[h]{c   c  c  c c c c c c c c c c c  }
      &\multicolumn{12}{c}{$N$} \\ \cmidrule{2-13}
       & 4 & 5 & 6 & 7 & 8 & 9 & 10 & 11 & 12 & 13 & 14 & 15 \\
      \hline
      $b$ &  |  &  | &   | &  17.81 &   44.09  &  82.45  &  127.59 &   167.71  &  219.89  &  264.01  & 298.47 & 341.50\\
      \hline
      $c_0$ & |  &  | &   | & -7.58 & -15.57 & -24.23 & -32.92 & -41.24 &  -48.94 &   -55.95 &   -62.25 &  -67.85 &  \\
      \bottomrule
    \end{tabular}
    \caption{Values of the fitted parameters $b$ and $c_0$, for spin $s=3$ and $N=7,8,\dots,15$.}
    \label{tables3}
\end{table}


These features, however, need a more careful analysis since it is extremely difficult and computationally challenging to study the data in the close vicinity of the critical value $r^*$, as the iterative procedure becomes extremely slow.
The best way to overcome this problem is to find clever ways to solve the TBA equations \eqref{eq:TBAeq} analytically, to have a more quantitative analysis of these models.
A first attempt in this direction is provided in \cref{appB}, with a simpler toy model.

\newpage
\setcounter{equation}{0}
\section{Conclusions}\label{sec5}
Following the inverse scattering program, we have constructed exact $S$-matrices of particles belonging
to spin $s$ multiplets of quantum $\uqsu2$ group for any half and integer spins.
The scalar factors in front of the $S$-matrices are derived exactly to satisfy the unitarity and crossing-symmetry in
\eqref{common}.
The first two simple cases $s=1/2$ and $1$ match exactly with known results of the quantum sine-Gordon and the sausage models.
We have found many new $S$-matrices in this way.
For example, explicit expressions of new $S$-matrix elements for $s=3/2$ are given in \eqref{exps32}.

We have derived the TBA equations which can be associated with graphs similar to but different from Dynkin diagrams.
Except for $s=1/2,1$, these graphs are different from those of classical Lie algebras of finite or affine types.
In our opinion, it is not a coincidence that the Hagedorn-like singularity occurs for non-Dynkin graphs for spin $s\ge 3/2$ and this fact deserves, with no doubt, further investigations.
We have shown how the pseudo-energies and the free energies become complex if the temperature scale becomes larger than the critical ones.
Although these exact results are based on a TBA with the same non-Dynkin graphs as shown in \cref{graph1,graph2} and a simplified toy kernel, we believe that the origin of the singularity should not be the kernel but the graphs which are of non-Dynkin types.
This is supported by the numerical analysis in which both the toy TBA and the TBA for $s\ge 3/2$ show qualitatively the same singularities.
Understanding this from mathematical considerations such as cluster algebras and hyperbolic algebras will be interesting.
In the context of integrable QFTs, the fact that the pseudo-energies become complex implies that the minimization of the free energy leading to the TBA equations fails and signals the presence of singularity.
Unfortunately, we have no theoretical tools beyond the critical scales.

The numerical data suggest an interesting universality in which the critical exponents are quite close to $0.5$
independently of the spin $s$ and the coupling constant $\gamma=1/N$.
If it is interpreted as $1/2$, the singular behavior is similar to a particular deformation by the energy-momentum $T\bar{T}$  where the free energy can be computed using the Burgers equations.
However, we want to point out that our results are qualitatively different from this $T\bar{T}$ deformation since the phase shifts from \eqref{common} are regular for all values of $s$ in the asymptotic limit $\theta\to\infty$ while that of the $T\bar{T}$ is singular.\footnote{Deformations by higher conserved charges $(T\bar{T})^{(s)}$ which give  CDD factors of the ``pole'' type can also lead to the Hagedorn singularity even though the phase shifts are regular as pointed out in \cite{camilo}.}

There are several unanswered questions, both conceptual and technical, which suggest some future investigations.
Considering that our $S$-matrices are based on the spin $s$ representations of $\uqsu2$,
we should understand why cases of $s=1/2,1$ are so special to have UV completeness while other values of $s$ cannot.
Also, we can ask if exact $S$-matrices based on other symmetry algebras than $\mathfrak{su}_2$ can show similar singularity.
Another interesting question is to study this singularity in the limit of $s\to\infty$ in the context of certain condensed matter problems like Haldane conjecture where the large spin limit plays a significant role \cite{Haldane}.
On the numerical side, a deeper understanding of the scaling function behavior in the vicinity of the critical point $r^*$, maybe using the techniques suggested in \cite{camilo} for the numerical integration of TBA, could shed more light on the phenomena appearing there.

\section*{Acknowledgements}
We thank Z. Bajnok, J. Balog, D. Bernard, O. Castro-Alvaredo, V. Dobrev, P. Dorey, B. Doyon, F. Essler, D. Fioravanti, S. Lukyanov, M. Mazzoni, S. Negro, R. Nepomechie,
L. Piroli, N. Reshetikhin, F. Sailis, M. Staudacher, S. van Tongeren, and G. Takacs for valuable discussions, useful comments, and suggestions. 
This work was supported by the National Research Foundation of Korea (NRF) grant
(NRF-2016R1D1A1B02007258) (CA),  and 
by Commission IV (Theory) of I.N.F.N. under the grants GAST and DOT4 (FR).

\appendix
\renewcommand{\theequation}{A.\arabic{equation}}
\setcounter{equation}{0}
\section{Explicit derivation of Bethe-Yang equations}\label{appA}
It is important to find explicit expressions of the kernels of TBA equations.
For this purpose, we introduce another set of functions
\bea
\mathsf{a}_n(\theta)&\equiv&\frac{i}{2\pi}\frac{d}{d\theta}\ln\mathsf{e}_n(\theta)=
\frac{\gamma}{\pi}\frac{\sin n\pi\gamma}{\cosh 2\gamma\theta-\cos n\pi\gamma},\\
\mathsf{b}_n(\theta)&\equiv&\frac{i}{2\pi}\frac{d}{d\theta}\ln\mathsf{g}_n(\theta)=
-\frac{\gamma}{\pi}\frac{\sin n\pi\gamma}{\cosh 2\gamma\theta+\cos n\pi\gamma}.
\eea
In terms of these functions, the kernels can be derived from (\ref{s00}-\ref{s0N}) and (\ref{snm}-\ref{sNN})
($n,m=1,\cdots,N-1$)
\bea
K_{0n}&=&K_{n0}=-\sum_{i=1}^{{\rm min}(n,2s)-1}\mathsf{a}_{|n-2s|+2i-1},\\
K_{0N}&=&K_{N0}=-\mathsf{b}_{2s},\\
K_{nm}&=&K_{mn}=\mathsf{a}_{|n-m|}+\mathsf{a}_{n+m}+2\sum_{i=1}^{{\rm min}(n,m)-1}\mathsf{a}_{|n-m|+2i},\\
K_{nN}&=&K_{Nn}=\mathsf{b}_{n-1}+\mathsf{b}_{n+1}.
\eea

It is more convenient to take Fourier transforms of the kernels.
With the convention 
\bea
\hat{f}(\omega)=\int_{-\infty}^{\infty}e^{i\omega\theta}f(\theta)d\theta,\quad
f(\theta)=\int_{-\infty}^{\infty}e^{-i\omega\theta}\hat{f}(\omega)\frac{d\omega}{2\pi},
\eea
the Fourier transforms of $\mathsf{a}_n(\theta)$ and $\mathsf{b}_n(\theta)$, with $\gamma=1/N$, are given by
\bea
\hat{\mathsf{a}}_n(\omega)=\frac{\sinh\left(\pi\frac{N-n}{2}\omega\right)}
{\sinh\left(\pi\frac{N}{2}\omega\right)},\qquad
\hat{\mathsf{b}}_n(\omega)=-\frac{\sinh\left(\pi\frac{n}{2}\omega\right)}
{\sinh\left(\pi\frac{N}{2}\omega\right)}.
\eea
From these, we can obtain the Fourier transforms of the kernels as follows:
\begin{align}
&\hat{K}_{00}=\frac{\sinh\left(\frac{N-2s}{2}\pi\omega\right)\sinh\left(s\pi\omega\right)}
{\sinh(\pi\omega)\,\sinh\left(\frac{N}{2}\pi\omega\right)},\label{Kh00}\\
&\hat{K}_{NN}=\frac{\sinh\left(\frac{N-2}{2}\pi\omega\right)}
{\sinh\left(\frac{N}{2}\pi\omega\right)},\label{KhNN}\\
&\hat{K}_{0N}=\hat{K}_{N0}=
\frac{\sinh\left(s\pi\omega\right)}{\sinh\left(\frac{N}{2}\pi\omega\right)},\label{Kh0N}\\
&\hat{K}_{N-1,N}=\hat{K}_{N,N-1}=-\frac{\sinh\left(\frac{N-2}{2}\pi\omega\right)}
{\sinh\left(\frac{N}{2}\pi\omega\right)},\label{KhNN1}\\
&\hat{K}_{0n}=\hat{K}_{n0}=
-\frac{\sinh\left(\frac{n}{2}\pi\omega\right)\sinh\left(\frac{N-2s}{2}\pi\omega\right)}
{\sinh\left(\frac{N}{2}\pi\omega\right)\sinh\left(\frac{1}{2}\pi\omega\right)},\qquad 1\le n<2s,\label{Kh0n1}\\
&\hat{K}_{0n}=\hat{K}_{n0}=
-\frac{\sinh\left(s\pi\omega\right)\sinh\left(\frac{N-n}{2}\pi\omega\right)}
{\sinh\left(\frac{N}{2}\pi\omega\right)\sinh\left(\frac{1}{2}\pi\omega\right)},
\qquad 2s\le n\le N-2,\label{Kh0n2}\\
&\hat{K}_{nN}=\hat{K}_{Nn}=
-\frac{2\sinh\left(\frac{n}{2}\pi\omega\right)\cosh\left(\frac{1}{2}\pi\omega\right)}
{\sinh\left(\frac{N}{2}\pi\omega\right)},\qquad 1\le n\le N-2,\label{KhnN}\\
&\hat{K}_{nm}=\hat{K}_{mn}=
\frac{\sinh\left(\frac{N-n}{2}\pi\omega\right)\sinh\left(\frac{m}{2}\pi\omega\right)
\sinh\omega}
{\sinh^2\left(\frac{1}{2}\pi\omega\right)\sinh\left(\frac{N}{2}\pi\omega\right)}-\delta_{nm},\qquad 
1\le m\le n\le N-1.\label{Khnm}
\end{align}
The kernel $\hat{K}_{00}$ can be obtained from \eqref{common}.
By defining the following kernels,
\bea
\hat{p}(\omega)&\equiv&\frac{1}{2\cosh\left(\frac{1}{2}\pi \omega\right)},\\
\hat{\mathcal{K}}_{nm}(\omega)&\equiv&\hat{K}_{nm}(\omega)+\delta_{nm},
\eea
one can easily check that the following functional relations are satisfied for all $1\le m\le N$,
\begin{align}
&\hat{\mathcal{K}}_{nm}=\delta_{nm}+\hat{p}(\eta_{n1}\hat{\mathcal{K}}_{n-1,m}
+\eta_{n,N-1}\hat{\mathcal{K}}_{n+1,m})-\hat{p}\delta_{n,N-2}\delta_{mN},
\quad 1\le n\le N-1,\label{KRnm}\\
&\hat{\mathcal{K}}_{0m}=-\hat{p}\hat{\mathcal{K}}_{2s,m}+\hat{p}\delta_{2s,N-1}\delta_{mN},
\label{KR0m}\\
&\hat{\mathcal{K}}_{Nm}=\delta_{Nm}-\hat{p}\hat{\mathcal{K}}_{N-2,m},\label{KRNm}\\
&\hat{\mathcal{K}}_{00}=1+\hat{p}^2\hat{\mathcal{K}}_{2s,2s},\label{KR00}
\end{align}
where we have used the short notation $\eta_{nm}=1-\delta_{nm}$, i.e.
\bea
\eta_{nm}=\begin{cases}
0,&n=m\\
1,&n\neq m.
\end{cases}
\eea
Inserting \eqref{KRnm} into \eqref{BetheYang} for $n=1,\cdots,N-1$, we can find
\bea
\hat{\sigma}_n+\hat{\tilde\sigma}_n=\hat{p}\left(\eta_{n1}\hat{\tilde\sigma}_{n-1}+
\eta_{n,N-1}\hat{\tilde\sigma}_{n+1}+\delta_{n,N-2}\hat{\sigma}_N+\delta_{n,2s}\hat{\sigma}_0\right)
\eea
From \eqref{KRNm} for $n=N$, we get
\bea
\hat{\sigma}_N+\hat{\tilde\sigma}_N=\hat{p}\hat{\tilde\sigma}_{N-2} + \delta_{N-1,2s}\hat{p}\hat{\sigma}_0,
\eea
and from \eqref{KR0m} and \eqref{KR00}, we get
\bea
\hat{\sigma}_0+\hat{\tilde\sigma}_0=\hat{q}+\hat{p}(\hat{\tilde\sigma}_{2s}-\delta_{2s,N-2}\hat{\sigma}_N),
\eea
where $\hat{q}$ is the formal Fourier transform of $\mathsf{m}\cosh\theta$.
Taking inverse Fourier transforms on these equations, we get
\begin{align}
&\sigma_{1}(\theta)+{\tilde\sigma}_1(\theta)=p\star{\tilde\sigma}_{2}(\theta) +\delta_{1,2s}p\star{\sigma}_{0}(\theta), \label{sigmaeq2}\\
&\sigma_{n}(\theta)+{\tilde\sigma}_n(\theta)=p\star{\tilde\sigma}_{n-1}(\theta) +p\star{\tilde\sigma}_{n+1}(\theta)+\nonumber\\
&\qquad\qquad\quad\qquad +\delta_{n,N-2}p\star{\sigma}_{N}(\theta)+\delta_{n,2s}p\star{\sigma}_{0}(\theta),\quad n=2,\dots,N-2, \label{sigmaeq3}\\
&\sigma_{N-1}(\theta)+{\tilde\sigma}_{N-1}(\theta)=p\star{\tilde\sigma}_{N-2}(\theta) +\delta_{N-1,2s}p\star{\sigma}_{0}(\theta),\label{sigmaeq4}\\
&\sigma_{N}(\theta)+{\tilde\sigma}_N(\theta)=p\star{\tilde\sigma}_{N-2}(\theta) + \delta_{N-1,2s}p\star \sigma_0(\theta),\label{sigmaeq5}\\
&\sigma_{0}(\theta)+{\tilde\sigma}_0(\theta)=\mathsf{m}\cosh\theta+p\star{\tilde\sigma}_{2s}(\theta) - \delta_{2s,N-2}p\star \sigma_N(\theta).\label{sigmaeq6}
\end{align}
Here we have used the universal kernel $p(\theta)$, defined in \cref{unikern}.
By minimizing the free energy with these constraints, we can express the TBA equations in terms of incidence matrix elements as \cref{eq:TBAeq}.


\renewcommand{\theequation}{B.\arabic{equation}}
\setcounter{equation}{0}
\newpage
\section{Simplified TBA equations with a toy kernel}\label{appB}
The TBA system \eqref{eq:TBAeq} is not solvable analytically and it is difficult to understand how the TBA develops the singularity.
To get an insight on this at a technical level, we consider the toy kernel\footnote{The normalization is chosen to match that of the integrated universal kernel from the previous sections.}
\bea
p(\theta)=\frac{1}{2}\delta(\theta).
\eea
The TBA equations become very simple, namely
\bea
\epsilon_n(\theta)=\delta_{n,0}r\cosh\theta-\frac{1}{2}\sum_{m=0}^{N}\mathbb{I}_{nm}
\log\left[1+e^{-\epsilon_m(\theta)}\right],\qquad n=0,1,\cdots,N,
\label{sTBA}
\eea
and can be expressed as a set of algebraic equations 
\bea
x_n(\theta)=e^{-r \cosh\theta\delta_{n,0}}\prod_{m=0}^{N}
\left[1+x_m(\theta)\right]^{\mathbb{I}_{nm}/2},\qquad x_n(\theta)\equiv e^{-\epsilon_n(\theta)},\quad
r\equiv \mathsf{m}R
\label{sYsys}
\eea
for each value of $\theta$. 
Since these are not integral equations anymore, we can analyze them analytically.
\begin{table}[h!]

\begin{center}
\begin{tabular}{|c |c |c |}
\hline
$k=2 s$& $N$ & $r^*$ \\ 
\hline
1& $\ge 4$& 0 \\  
\hline
2& $\ge 5$& $< 10^{-4}$  \\
\hline
3&4 & $\log 4=1.38629436...$  \\
3&5 & 1.38629436  \\
3&6 & 1.38629436  \\
3&7 & 1.38629436  \\
3&8 & 1.38629436  \\
\hline
4&5 & $\log 8=2.07944154...$  \\
4&6 & 2.07944154  \\
4&7 & 2.07944154  \\
4&8 & 2.07944154  \\
\hline
5&7 & 2.57478171 \\
5&8 & 2.57478171  \\
\hline
\end{tabular}

\caption{$r^*$ for various $s$ and $N$. We observe that $r^*$ depends only on $s$ and is independent of $N$. For $s=1/2,1$, $r^*$ is zero, meaning that there is no singularity}.
\label{numvalues}
\end{center}

\end{table}

These TBA equations 
whose graph is shown above in \cref{graph2}
are exactly solvable by Mathematica for $N=2s+1$ with $2s=3,4$
\footnote{Exact solutions for higher values of $s$ are beyond the capacity of our Mathematica code.} 
in terms of $a=e^{-r\cosh\theta}$,
\bea
x_0^{[2s=3]}&=&-\frac{5}{2}+\frac{1}{2a^2}-a-\frac{(1+a)^2}{2a^2}\sqrt{1-4a},\\
x_0^{[2s=4]}&=&-\frac{3}{2}+\frac{1}{2a}-a-\frac{(1+a)}{2a}\sqrt{1-8a}.
\eea
All other exact expressions for $x_n$'s are also found but we will not put them here since they are much more complicated and not relevant in further discussions.

These results show that the pseudo-energies can be real if 
\bea
e^{-r\cosh\theta}\le\frac{1}{4}\quad{\rm for}\ 2s=3;\qquad
e^{-r\cosh\theta}\le\frac{1}{8}\quad{\rm for}\ 2s=4,
\eea
for any value of $\theta$. 
Therefore, the critical values $r^*$, which are the maximum values of $r$ for them to remain all real, 
 are found by considering $\theta=0$, namely,
\bea
r^*=\begin{cases}
\log 4,&{\rm for}\ 2s=3\\
\log 8,&{\rm for}\ 2s=4.
\end{cases}
\eea

For other values of $s$ and $N$, we can solve only numerically to find $r^*$.
We list them for different values of $N$ in \cref{numvalues}. 
It is interesting to notice that the critical values $r^*$ where the solutions turn into complex numbers, depend only on the spin $s$ and not on the coupling constant $N=1/\gamma$. 
We do not understand this analytically but it is definitely due to the exceptionally simplified kernel.
For $s=1/2,1$ where no singularity occurs, we observe that the solutions are real for all $r$ as expected.

\begin{figure}
\begin{minipage}{0.5\textwidth}
    \centering
    \includegraphics[width=\linewidth]{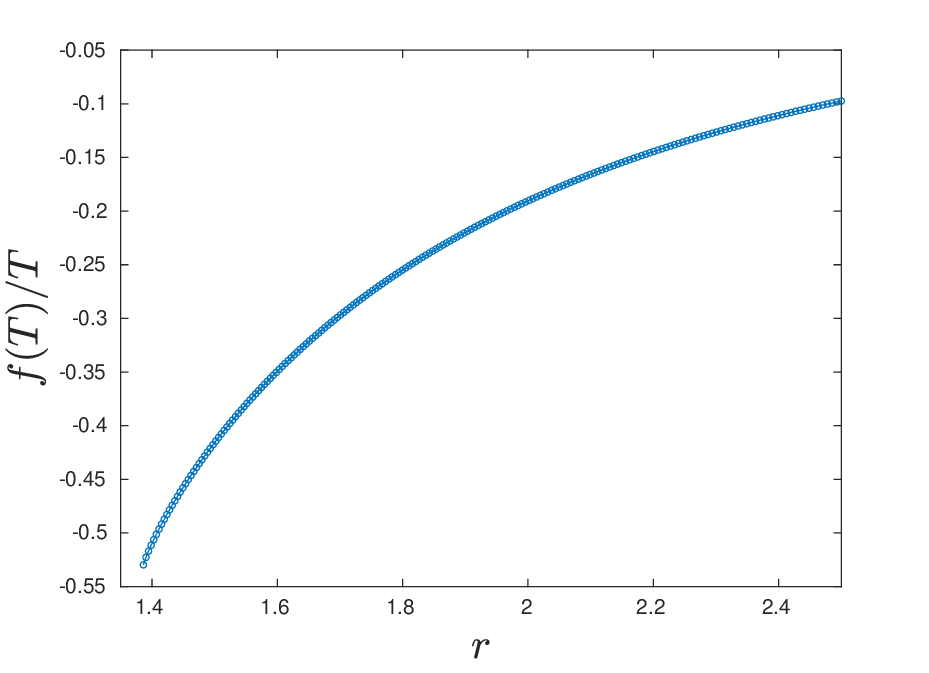}
  \end{minipage}
  \hfill
  \begin{minipage}{0.5\textwidth}
    \centering
    \includegraphics[width=\linewidth]{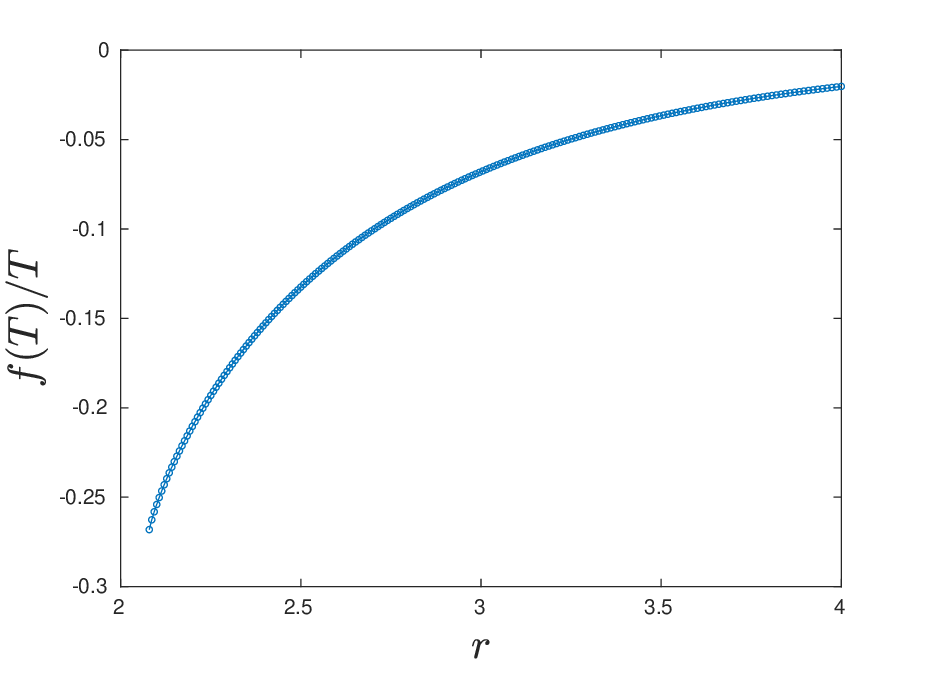}
  \end{minipage}
\caption{Toy TBA with $s=3/2, N=4$ (left) and $s=2, N=5$ (right).}
\label{fig9}
\end{figure}

The above exact solutions of the simplified TBA can be used to analyze the free energy
using \eqref{freen}. For $2s=3$, one finds
\bea
\frac{f(T)}{T}=-\int_{0}^{\infty}\frac{\mathsf{m}}{\pi}\cosh\theta\,\ln\left[
-\frac{3}{2}+\frac{1}{2a^2}-a-\frac{(1+a)^2}{2a^2}\sqrt{1-4a}\right]d\theta,\quad a=e^{-r\cosh\theta},
\eea
Although this expression is given in terms of relatively simple integrals, it can not be expressed analytically.
Instead, we perform this numerically and plot the free energy as a function of $r$ near $r^*$ in \cref{fig9} which show qualitatively similar behaviors as \cref{fig5}.

\begin{thebibliography}{99}
\bibitem{zamzam} A.B. Zamolodchikov and Al.B. Zamolodchikov, {\em Factorized S-matrices in two dimensions as the exact solutions of certain relativistic quantum field models}, 
Annals Phys. {\bf 120} (1979) 253
\bibitem{BerLeC} D. Bernard and A. LeClair, {\em Residual quantum symmetries of the restricted sine-Gordon theories}, Nucl. Phys. {\bf B340} (1990) 721
\bibitem{alyosha_tba} Al.B. Zamolodchikov, {\em Thermodynamic Bethe ansatz in relativistic models. Scaling three state Potts and Lee-Yang models}, Nucl. Phys. {\bf B342} (1990) 695
\bibitem{foz} V.A. Fateev, E. Onofri and Al.B. Zamolodchikov, {\em Integrable deformations of the O(3) sigma model. The sausage model}, Nucl. Phys. {\bf B406} (1993) 521
\bibitem{ads/cft} N.~Beisert, C.~Ahn, L.F.~Alday, Z.~Bajnok, J.M.~Drummond, L.~Freyhult, N.~Gromov, R.A.~Janik {\it et al.}, {\em Review of AdS/CFT Integrability: An Overview},  Lett.\ Math.\ Phys.\  {\bf 99} (2012) 3, arXiv:1012.3982 [hep-th]
\bibitem{Hagedorn} R. Hagedorn, {\em Statistical thermodynamics of strong interactions at high energies}, Nuovo Cim. Suppl. {\bf 3} (1965) 147
\bibitem{stringHag} J. Atick and E. Witten,  {\em The Hagedorn transition and the number of degrees of freedom of string theory}, Nucl. Phys. {\bf B310}  (1988) 291
\bibitem{ttbar1} 
  F.A. Smirnov and A.B. Zamolodchikov,  {\em On the space of integrable quantum field theories}, Nucl. Phys. {\bf  B915}  (2017) 363, arXiv:1608.05499 [hep-th]
\bibitem{ttbar2} 
  A. Cavagli\`a, S. Negro, I.M. Szecsenyi and R. Tateo, {\em $T{\bar T}$-deformed 2D quantum field theories}, JHEP {\bf 10}  (2016) 112, arXiv:1608.05534 [hep-th]
  \bibitem{ttbar3} 
  S. Dubovsky, V. Gorbenko and M. Mirbabayi, {\em Asymptotic fragility, near AdS2 holography and $T{\bar T}$},  JHEP {\bf 09}  (2017) 136, arXiv:1706.06604 [hep-th]
\bibitem{ttbar4} 
  S. Dubovsky, V. Gorbenko,  and C. Hernandez-Chifflet, {\em $T{\bar T}$ partition function from topological gravity}, JHEP {\bf  09}  (2018) 158, arXiv:1805.07386 [hep-th]
\bibitem{ttbar5} 
G. Mussardo and P. Simon, {\em Bosonic-type S-Matrix, Vacuum Instability and CDD Ambiguities}, Nucl. Phys. {\bf B578} (2000) 527, arXiv:hep-th/9903072
\bibitem{AhnLeC} C. Ahn and A. LeClair, {\em On the classification of UV completions of integrable $ T{\bar T}$  deformations of CFT}, JHEP {\bf 2022} (2022) 179, arXiv:2205.10905 [hep-th]
\bibitem{AladimMartins} S.R. Aladim and M.J. Martins, {\em Bethe ansatz and thermodynamics of a $SU(2)_k$ factorizable S matrix}, Phys. Lett. {\bf B329} (1994) 271
\bibitem{Kir-Resh1} A. Kirillov and N.Y. Reshetikhin, {\em Exact solution of the integrable XXZ Heisenberg model with arbitrary spin. I. The ground state and the excitation spectrum}, J. Phys. {\bf A20} (1987) 1565
\bibitem{Kir-Resh2} A. Kirillov and N.Y. Reshetikhin, {\em Exact solution of the integrable XXZ Heisenberg model with arbitrary spin. II. Thermodynamics of the system}, J. Phys. {\bf A20} (1987) 1587
\bibitem{Kirillov} A. Kirillov, {\em Clebsch-Gordan quantum coefficients}, 	J. Sov. Math. {\bf 53} (1991) 264
\bibitem{Lusztig} G. Lusztig,  {\em Modular representations and quantum groups}, Contemp. Math. {\bf 82} (1989) 59
\bibitem{Hou} Bo-Yu Hou, Bo-Yuan Hou and Z.-Q. Ma, {\em Quantum Clebsch-Gordan coefficients for nongeneric q values}, J. Phys.  {\bf A25} (1992) 1211
\bibitem{Ruegg} H. Ruegg, {\em A Simple Derivation of the Quantum Clebsch Gordan Coefficients for $SU(2)_q$}, J. Math. Phys. {\bf 31} (1990) 1085
\bibitem{TakSuz} M. Takahashi and M. Suzuki, {\em One-Dimensional Anisotropic Heisenberg Model at Finite Temperatures}, Prog. Theor. Phys. {\bf 48} (1972) 2187
\bibitem{alyosha_tba_rsos}Al.B. Zamolodchikov, {\em Thermodynamic Bethe ansatz for RSOS scattering theories}, Nucl. Phys. {\bf B358} (1991) 497
\bibitem{ingo17} L. Hilfiker and I. Runkel, {\em Existence and uniqueness of solutions to Y-systems and TBA equations}, Annales Henri Poincare {\bf 21} (2019) 941, arXiv:1708.00001 [math-phys]
\bibitem{sundb} B. Sundborg, {\em The Hagedorn transition, deconfinement and $\mathcal{N}=4$ SYM theory}, Nucl. Phys. {\bf B573} (2000) 349, arXiv:hep-th/9908001 [hep-th]
\bibitem{Haldane} F. Haldane, {\em Continuum dynamics of the 1-D Heisenberg antiferromagnet: Identification with the O(3) nonlinear sigma model}, Phys. Lett. {\bf A93} (1983) 464
\bibitem{camilo} G. Camilo, T. Fleury, M. Lencs\'es, S. Negro and A.B. Zamolodchikov, {\em On factorizable S-matrices, generalized $T\bar{T}$ and the Hagedorn transition}, JHEP {\bf 2021} (2021) 62, arXiv:2106.11999 [hep-th]


\end{thebibliography}
\end{document}